\documentclass[12pt]{article}
\input epsf.tex

\def\ie{{\it i.e.}}
\def\eg{{\it e.g.}}

\let\badcite=\cite
\def\cite{~\badcite}

\def\slashchar#1{\setbox0=\hbox{$#1$}           
   \dimen0=\wd0                                 
   \setbox1=\hbox{/} \dimen1=\wd1               
   \ifdim\dimen0>\dimen1                        
      \rlap{\hbox to \dimen0{\hfil/\hfil}}      
      #1                                        
   \else                                        
      \rlap{\hbox to \dimen1{\hfil$#1$\hfil}}   
      /                                         
   \fi} 
    \def\slashword#1{\setbox0=\hbox{$#1$}        
  \dimen0=\wd0                                   
   \setbox1=\hbox{/} \dimen1=\wd1                
   \ifdim\dimen0>\dimen1                         
      \rlap{\hbox to \dimen0{\hfil\bf---\hfil}} %
      #1                                         %
   \else                                         
      \rlap{\hbox to \dimen1{\hfil$#1$\hfil}}    
      /                                          
    \fi}                                         %

\catcode`@=11
\newdimen\vbigd@men                             

\def\vbig#1#2{{\vbigd@men=#2\divide\vbigd@men by 2%
   \hbox{$\left#1\vbox to \vbigd@men{}\right.\n@space$}}}
\catcode`@=12

\catcode`@=11
\def\citenum#1{\csname b@#1\endcsname}
\catcode`@=12

\def\dofig#1#2{\centerline{\epsfxsize=#1\epsfbox{#2}}}

\begin{document}
\begin{titlepage}

\begin{flushright}
{SCUPHY-TH-05001}\\
{SHEP-04-18}\\
{TUHEP-TH-04146}\\   
\end{flushright}

\vskip -0.6cm
\bigskip


{\Large{\bf
\centerline{Pair-produced heavy particle topologies:}
\centerline{MSSM neutralino properties}
\centerline{at the LHC from gluino/squark cascade decays}
}}


\renewcommand{\thefootnote}{\fnsymbol{footnote}}

\bigskip
\centerline{\bf M. Bisset, N. Kersting\footnote{present address: 
Department of Physics, Sichuan University}, J. Li}
\centerline{{\it Center for High Energy Physics and Department of
Physics,}}
\centerline{{\it Tsinghua University,}}
\centerline{{\it Beijing, 100084 P.R. China}}
\vskip 0.25cm
\centerline{\bf F. Moortgat}
\centerline{{\it PH Department, CERN, CH-1211 Geneva 23, Switzerland}}
\vskip 0.25cm
\centerline{\bf S. Moretti}
\centerline{{\it School of Physics and Astronomy}}
\centerline{{\it University of Southampton}}
\centerline{{\it Highfield, Southampton SO17 1BJ, UK}}
\vskip 0.25cm
\centerline{\bf Q.L. Xie}
\centerline{{\it Department of Physics}}
\centerline{{\it Sichuan University}}
\centerline{{\it Chengdu, 610065 P.R. China}}

\renewcommand{\thefootnote}{\arabic{footnote}}
\setcounter{footnote}{0}

\vskip -0.3cm
\bigskip

\begin{abstract}

Processes of the form 
$pp\to anything \to X_i X_j \to x\bar{x} + y\bar{y}
\; (+ \slashchar{E})$ are studied via a technique that may be 
viewed as an adaptation of time-honoured Dalitz plot analyses.
$X_i$ and $X_j$ are new heavy states (with $i,j=1...n$),
which may be identical or distinct; and $x\bar{x}$ and $y\bar{y}$ 
are necessarily distinct Standard Model (SM) fermion pairs whose
invariant masses can be measured.
A Dalitz-like plot of said invariant masses, $M(x\bar{x})$
{\it vs.$\!\!\!$ } $M(y\bar{y})$, exhibits a topology 
connected to the masses and specific decay chains of $X_i$ and $X_j$. 
Aside from relatively minor details, observed patterns
consist of a collection of box and wedge shapes.  This collection
is model-dependent:  comparison of the observed pattern to 
the possibilities for a specific model yields information
on which new particle pair combinations are actually being produced,
information beyond that extractable from conventional one-dimensional 
invariant mass distributions.  
The technique is illustrated via application to the
Minimal Supersymmetric Standard Model (MSSM) process
$pp \to \tilde{g}\tilde{g},\tilde{g}\tilde{q},\tilde{q}\tilde{q} 
\to \widetilde{\chi}_i^0 \widetilde{\chi}_j^0
\to e^+ e^- \, + \, \mu^+ \mu^- \; (+ \slashchar{E})$. 
Here the heavy states are neutralinos $\widetilde{\chi}_i^0$ ($i=2,3,4$) 
--- note $\widetilde{\chi}_1^0$ is excluded ---
which are produced in gluino/squark ($\tilde{g}$/$\tilde{q}$) 
cascade decay chains.  
Even with fairly modest expectations for the LHC performance during 
the first few years, this method still provides substantial insight into
the neutralino mass spectrum and couplings 
if gluino/squark masses are relatively low ($\simeq \, 400\, \hbox{GeV}$).


\end{abstract}

\newpage
\pagestyle{empty}

\end{titlepage}

\section*{Introduction}
\label{sec:intro}

The field of particle physics is nearing a critical juncture: 
up to now the highly successful SM --- whose
predictions of various cross-sections and precision observables are in 
excellent agreement with data from the most advanced particle accelerators
to date --- has been sufficient to meet the experimental demands; however, 
the SM is theoretically incomplete and cannot continue to describe physics 
at energies much higher than $1\, \hbox{TeV}$.  
Most theoretical extensions of the SM designed to address this problem 
predict new heavy degrees of freedom at or near the TeV-scale.
The soon-to-be-completed LHC, 
with a centre-of-mass energy of $14\, \hbox{TeV}$, 
should readily produce such heavy particles if they couple significantly 
to SM ones.  Then experiment will certainly require more guidance 
than the SM can provide.  Different extensions to the SM 
differ in the predicted number and types of new heavy particles.
It is therefore imperative to understand what the decays of such  
heavy particles (at least some of which are typically unstable) would 
look like at a hadron collider in as model-independent a way as possible.

This work is particularly concerned with neutral heavy particles
produced in pairs, $X_i X_j$, 
with $i,j=1...n$ --- the exact value of $n$ being model dependent, but for
the present work it may be any integer greater than 1
(thus $X_i$ and $X_j$ may or may not be distinct).
These pairs may be produced directly as per $pp \to X_i X_j$ 
or as a result of cascade decays from the production
of other even heavier new particles (in fact the latter production mode is
dominant in the specific case examined below).

The introduction of new heavy particle states often comes with the
introduction of a new conserved quantum number (or numbers) associated 
with a new discrete ${\cal Z}_n$ symmetry (or symmetries) --- for 
example, the attractive $R$-parity $\!\!\!\!\!$ \cite{GenRev}
conservation in numerous SUSY extensions\footnote{Henceforth the acronym
`SUSY' will be used for both `supersymmetry' and `supersymmetric'.} 
to the SM. 
Another example is found in some little Higgs models in which an extra 
${\cal Z}_2$ symmetry is introduced to tame excessive flavour-changing 
neutral
processes $\!\!\!\!\!$ \cite{Arkani-Hamed:2002pa}.  Conservation of such a 
new quantum number(s) typically dictates the pair production of new 
particle states as well as the stability of the lightest new particle 
which is ``odd'' under the new symmetry (features typically 
associated with $R$-parity conserving SUSY scenarios but in fact they 
seem to be more generally applicable).

Any sample of events collected over time may 
be a superposition of different $X_i X_j$ channels.
The technique introduced here is ideally suited for
precisely this situation.
Unlike at an $e^+e^-$ collider, at a hadron collider
the centre-of-mass energy of the parton-level hard scattering process 
cannot be controlled, and thus said parton-level centre-of-mass energy 
cannot be incrementally raised to scan through the different  
$X_i X_j$ thresholds.  Rather, all such channels may be produced 
simultaneously and must subsequently be disentangled to
the extent it is possible in the decay analysis.
 
The new heavy particles are assumed to decay (possibly indirectly) 
into pairs of SM fermions $X_i \to y\bar{y}$ (accompanied in some models 
containing stable but undetectable heavy particle states by the observation 
of substantial missing energy in the detector).  
Thus pair production and subsequent decay of the 
new heavy particles can result in end-states of the form $x\bar{x} \; + \; 
y\bar{y}$ , 
where $x\bar{x}$ and $y\bar{y}$ are {\em distinct} SM 
particle
pairs whose invariant masses are measurable
with sufficient precision. For example, same-flavour
oppositely-charged lepton pairs $e^+ e^-$ and $\mu^+ \mu^-$ 
(utilised in the following application to the MSSM $\!\!\!\!\!$
\cite{GenRev}) might be chosen since these are most easily extracted from
the overwhelming QCD backgrounds at a hadron collider. 
Other choices, such as $b \overline{b}$ or $\tau^- \tau^+$, are also
possible though, and might prove more appropriate in some cases.
Decays to SM gauge bosons may merit attention, though with decays to
$Z^0$'s a $Z^0$-veto to reduce backgrounds is no longer possible 
while decays to $W$'s will require reconstruction of hadronically-decaying
$W$'s.  The remainder of this work concentrates on decays into
pairs of fermions, and, more specifically, into electrons and muons.
Use of similar 
ideas for the pair-production of charged states ($X_i^+ X_j^-$) also
might merit future investigation. 

Speaking generally, the experimentally measurable quantities of interest 
are the fermion pair invariant masses $M(x\bar{x})$ and $M(y\bar{y})$. 
Other processes besides the sought-after heavy particle decays may
also produce an $(x\bar{x},y\bar{y})$-topology.
Thus cuts will probably be needed to purify the event sample, and a 
partially-contaminated sample may have to suffice.  
It will be shown below that, for several realistic MSSM scenarios including
both signals and backgrounds, making a two-dimensional 
Dalitz-like $\!\!\!$ \cite{Dalitzetc} plot of
$M(x\bar{x})$ vs. $M(y\bar{y})$ can reveal information about the spectrum 
of the heavy particles produced (kinematics) as well as relative production 
cross sections (dynamics).

\section*{Topological Analysis}

Any Dalitz-like plot of $M(x\bar{x})$ {\it vs.} $M(y\bar{y})$ resulting
from 
heavy particle pair production will be a superposition of specific 
topological shapes.  At the coarsest level, these shapes may
be bifurcated into two types:
\begin{itemize}
\item{\bf Box-like}
--- A `box' in the $M(x\bar{x})$-$M(y\bar{y})$ plane results from the
decay  
\begin{equation}
X_i X_i \to x\bar{x} + y\bar{y} \;\; (+ \slashchar{E}) \; ,
\end{equation} 
since the invariant masses $M(x\bar{x})$ and $M(y\bar{y})$ are bounded 
from below by the masses of $x,y$ (approximately zero if $x,y$ are leptons) 
and above by the maximum for $M_{X_i}-\slashchar{E}$ (this is a well-defined 
limit if the model in question completely accounts for $\slashchar{E}$ by 
particles which do not decay in the detector). 

\item{\bf Wedge-like} 
--- A `wedge' or `L-shape' results from the decay  
\begin{equation}
X_i X_j \to x\bar{x} + y\bar{y} \;\; (+ \slashchar{E}) \;\;\;\;  
(i\ne j) \; ;
\end{equation}
{\ie}, if the $x\bar{x}$ came from $X_i$ then the $y\bar{y}$ 
presumably comes from $X_j$ --- here it is assumed neither 
$x$- nor $y$-flavour number is violated in the heavy-particle 
decays\footnote{It is possible, for example, to 
have $e^{\pm}{\mu}^{\mp}$ decays from neutralinos in the 
lepton-flavour-conserving MSSM, but the branching ratios (BRs) 
for such decay modes are small and generally negligible.}. 
Therefore $0 < M(x\bar{x}) < M_{X_i} - \slashchar{E}  $ 
and $ 0 < M(y\bar{y}) < M_{X_j} -  \slashchar{E} $. 
On the other hand if the decays are swapped then 
$0 < M(x\bar{x}) < M_{X_j} -  \slashchar{E} $ 
and $ 0 < M(y\bar{y}) < M_{X_i} -  \slashchar{E}$; 
the superposition of these two strips forms the wedge.

\end{itemize}

The manner in which $X_i$ and $X_j$ decay may introduce new features on 
top of these two basic forms.  For example, whether the decay proceeds
through a series of two-body decays or via a three-body decay.
Furthermore, if some $X_i$ involved in the decay chain has
two or more ways to decay to $x\bar{x}$ and $y\bar{y}$; {\eg}, 
if two or more decay chains resulting in 
$X_i \to X_j + x\bar{x} (\hbox{or}~y\bar{y})$  are
kinematically allowed for any given $i$,
Dalitz-like plots will have `stripes'
extending from each of the endpoints of these decays to zero 
(or to $m_{x,y}$ if these are not approximately massless); these 
stripes will overlay the basic box/wedge structure outlined above.

If the types of decay chains the $X_i$ follow are known and
in particular if one type dominates ({\eg}, two-body decays
through one or more known intermediate states), the
shape of the $x\bar{x}$($y\bar{y}$) invariant mass spectrum 
can be predicted and
this information used to compare densities of points in different
regions of the Dalitz-like plot; this in turn allows one to measure
ratios of cross section $\times$ BR for the different modes $X_iX_j$
which are responsible for the various Dalitz shapes.
The Dalitz-like plots can then provide information about
{\it dynamics} in addition to {\it kinematics} (contained in the
location of the endpoints).

In any particular model there will be a set number of heavy particles 
expected to be produced at LHC energies; therefore the types of possible 
boxes and wedges is likewise set and the number of possible box-wedge 
combinations (with possible overlaying stripes) is fixed. 
Only some of these combinations are topologically distinct. 
For example, consider a sample of events where two $X_i X_i$-type
production modes dominate.  This will yield a Dalitz-like plot
that looks like a `box within a box' (ignore stripes for the
moment).  The topology alone would not indicate whether 
$X_2 X_2$ and $X_3 X_3$ are being produced or
$X_4 X_4$ and $X_5 X_5$ are being produced.
A collection of Dalitz patterns form a topological class if they can be
transformed into each other by any amount of dilation; {\ie},
they can be deformed into each other without crossing any kinematical 
hard edges. 
Furthermore, a wedge of type {\it ij} is difficult to distinguish from 
an {\it ij}-wedge and an {\it ii}-box ($M_{X_i} < M_{X_j}$) combination
and thus these two cases will be treated as topologically equivalent.  
If stripes are present the amount of degeneracy escalates 
(see subsequent application to SUSY).

\section*{Application to Sparticle Decays}

The $R$-parity conserving MSSM is next considered as a test-case for this
technique. 
In the MSSM, there are four distinct 
neutralinos\footnote{In what follows, we will often refer to neutralinos
collectively by the shorthand ``--inos''.}, the lightest of which, 
$\widetilde{\chi}_1^0$, is supposed to be stable and undetectable
({\it e.g.}, in minimal supergravity-inspired models).  
These --inos are the physical eigenstates resulting from the two pairs 
of neutral electroweak (EW) gauginos and Higgsinos in the MSSM.
Consider --ino pair production.
More specifically, the modes of interest here are pair production 
of {\em heavy} --inos:
\begin{equation}
pp \to \hbox{intermediate(s)} 
\to \widetilde{\chi}_i^0 \widetilde{\chi}_j^0
\;\;\;\; (i,j=2,3,4),
\label{inopairprod}
\end{equation} 
where {\em both} of the --inos subsequently decay (in the detector,
as expected in all MSSM scenarios)
leading to final states of the type described above.  Thus neither
--ino is allowed to be the stable Lightest Supersymmetric Particle (LSP)
$\widetilde{\chi}_1^0$.
However, in each event two LSP $\widetilde{\chi}_1^0$'s are  
subsequently produced from the decays of the two initial heavy
--inos, possibly through a chain of decays, along with the
$e^+e^-$ and ${\mu}^+{\mu}^-$ pairs we demand\footnote{Other
SM fermions aside from isolated electrons/muons may also
be present in the final state.}.
\begin{figure}[t!]
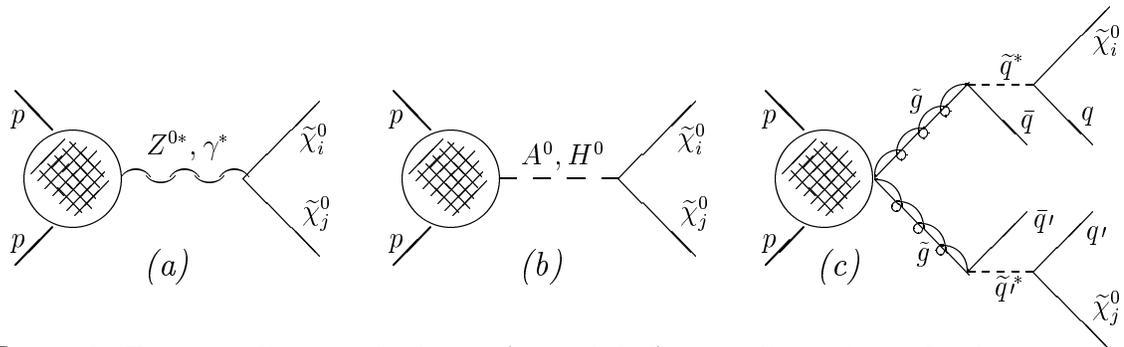

\vskip 1.2cm
\hspace{0.5cm}\dofig{2.40in}{figure1.eps}
\vskip -6.0cm
\caption{{\small
Feynman diagrams for heavy ($i,j = 2,3,4$) neutralino pair production
\newline
mechanisms:
{\it (a)} `direct' production via EW gauge boson;
{\it (b)} Higgs-mediated production; and
{\it (c)} production via cascade decays of gluinos (shown here)
or via gluino/squark or squarks --- to obtain these diagrams, make one
or both of the squarks in $(c)$ on-mass shell and remove the associated
gluino(s) and the connected quark(s).
}}
\label{inoinoprod}
\end{figure}
At the LHC, heavy --ino pair production occurs via
virtual SM gauge bosons (termed ``direct production''),
via the decays of heavy Higgs bosons, or via cascade decays 
of coloured squarks and gluinos (see Figure \ref{inoinoprod}).
The last of these, production via coloured intermediates,
\begin{equation}
pp \to \tilde{g} \tilde{g}, \, \tilde{g}\tilde{q}, \, \tilde{q}\tilde{q}
\to \widetilde{\chi}_i^0
\widetilde{\chi}_j^0 \;\;\;\; (i,j=2,3,4),
\label{glupairprod}
\end{equation}
will be the focus of the current work\footnote{Here taken to 
include $\tilde{g}\widetilde{\chi}$,$\tilde{q}\widetilde{\chi}$ 
production modes, which in fact only make minor contributions.}.
Due to the strong coupling,
(\ref{glupairprod}) has the potential for yielding the largest
number of signal events, if the intermediate gluinos and squarks are
sufficiently light.  Rates from EW direct production are typically too 
low for the technique described herein to be effectively utilised, except
perhaps in certain minor regions of un-excluded parameter space or
with allowance for ample time to gather more events
$\!\!\!\!\!$ \cite{inPrep}.
Since the lightest MSSM Higgs boson ($h^0$) can only yield 
LSP-containing --ino pairs, the Higgs-mediated production modes
of interest involve the heavier MSSM Higgs bosons 
($H^0$ and $A^0$).  Their masses need to be in the correct range 
to get sufficient Higgs boson production and yet have open decay modes to 
exclusively-heavy --ino pairs.  This is certainly possible,
as will be documented in another work $\!\!\!\!\!$ \cite{LH2003,inPrep}.  
However, the EW production rate will lead to a smaller number of potential
signal events than for (\ref{glupairprod}).
Thus, (\ref{glupairprod}) should be the main
source of --inos at the LHC
if gluinos are light ($\sim 400\, \hbox{GeV}$).  In the
current work, inputs for gluinos and squarks will be
set near the lower end of their allowed mass ranges while the
input Higgs boson mass will be fixed fairly high up
(${\sim}700\, \hbox{GeV}$)\footnote{These
restrictions are reversed in a detailed look at the decays of heavy MSSM
Higgs bosons into --inos in $\!\!\!\!$ \cite{LH2003,inPrep}.}.

Aside from the larger possible signal rates with (\ref{glupairprod})
as compared to with the two EW production mechanisms, there are   
two other seminal distinctions between (\ref{glupairprod}) and the other
two that can strongly influence the analysis.
First, as a side-product to producing --inos, the decaying gluinos/squarks
in (\ref{glupairprod}) also typically lead to jet activity in the final
state, whereas the other two production mechanisms may be hadronically quiet 
much or at least some of the time.  Thus backgrounds to 
(\ref{glupairprod}) may be more severe / less amenable to cuts.  This
could bring the signal rate {\em after cuts} down to the level of the
other two processes.  In fact, it will be shown in the simulations section
to follow that the backgrounds are not so severe.  Further, demanding the 
presence of jets is actually useful in reducing some backgrounds.

The second point warranting attention is that in (\ref{glupairprod}) the
two --inos are produced separately, whereas in the two EW processes there
is an --ino--ino$^{(\prime )}$-$S$ vertex (where $S$ is a SM gauge boson
or an MSSM heavy Higgs boson).  If the cascade decays were solely from
$\tilde{g}\tilde{g}$ and/or $\tilde{q}\tilde{q}$ production,
where here both squarks are of the same flavor and had the same
$SU(2)_{\hbox{\smash{\lower 0.25ex \hbox{${\scriptstyle L}$}}}}$
quantum number ({\it e.g.}, $\widetilde{u}_L \widetilde{u}_L^*$
$\widetilde{d}_R \widetilde{d}_R^*$, {\it etc.}), this would {\em reduce}
the number of possible topologies that can result from (\ref{glupairprod})
relative to the other production mechanisms
(that is, considering $\widetilde{\chi}_i^0\widetilde{\chi}_i^0$,
$\widetilde{\chi}_j^0\widetilde{\chi}_j^0$ and
$\widetilde{\chi}_i^0\widetilde{\chi}_j^0$ production with $i \ne j$,
knowing two of the three rates would determine the remaining one).
However, $\tilde{g}\tilde{q}$ production is very 
significant\footnote{Further, squark-initiated processes are likely
to contain extra jets which can increase the percentage of these
events that will pass a cut on the minimum number of jets in an event 
that will be imposed.}
(in fact, when all $\tilde{g}\tilde{q}$ combinations are added together, 
their combined rate is larger than either the $\tilde{g}\tilde{g}$ rate or
the combined $\tilde{q}\tilde{q}^{(\prime )}$ 
production rate $\!\!\!\!$ \cite{NPB_BHSZ}).  If all
the different squarks always decayed into gluinos, the afore-mentioned 
reduction in possible topologies would still occur.  
Actually, for the MSSM parameters herein considered, the different 
squarks decay into gluinos with BRs ranging from ${\sim}40$\% to
${\sim}95$\% (save for stops, which cannot decay into gluinos and top
quarks in the cases examined), and the remaining times decay directly
into charginos and neutralinos with differing BRs into the individual 
--inos, which would tend to restore the more general range of
topologies {\em if} the different squark flavours contributed comparably.
However, this is not the case --- contributions from the 
$\widetilde{u}_L$ squark are fairly dominant, and the BRs for this squark
tend not to differ markedly from those of the gluino.  So the reduction in 
topologies is partially true.  How much this is so will be quantified
later when specific points in the MSSM parameter space are discussed
(see Tables \ref{tab:branchings} and \ref{tab:percents}).

\begin{figure}[t!]
\begin{minipage}[t]{2.75in}
\dofig{2.75in}{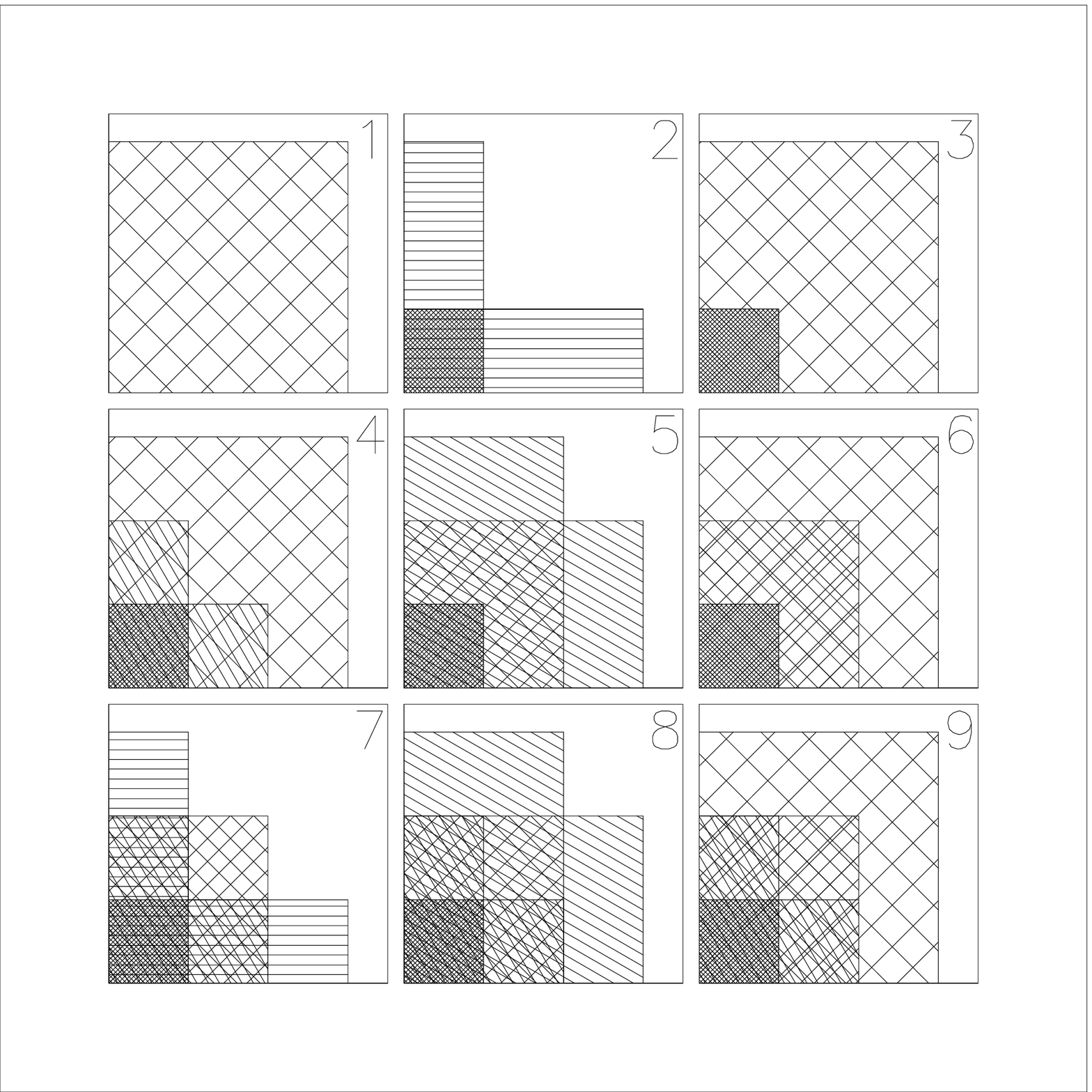}
\end{minipage}
\begin{minipage}[t]{2.75in}
\dofig{2.75in}{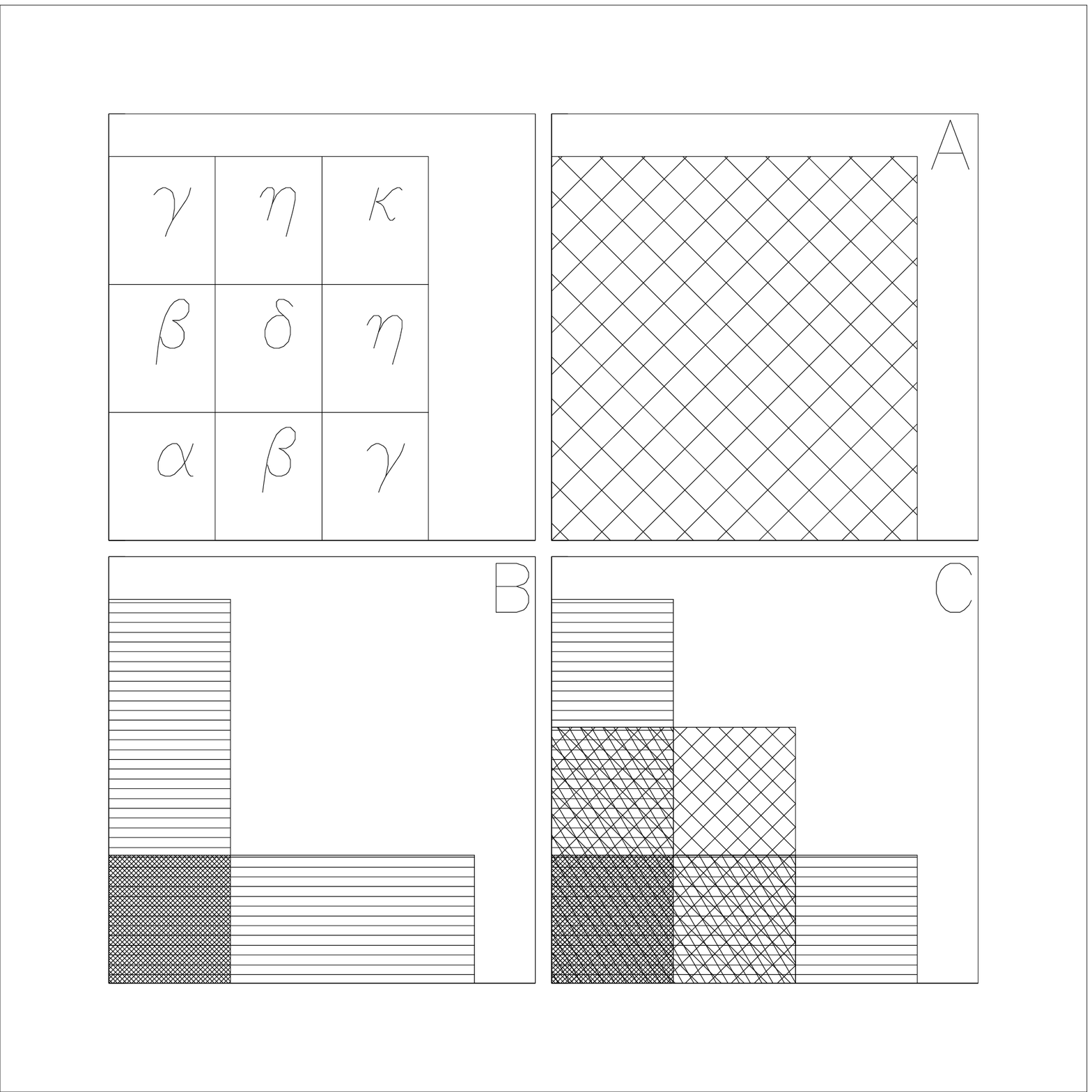}
\end{minipage} \hfill
\caption{{\small
At left, all possible distinct Dalitz patterns initiated 
by production of $\widetilde{\chi}_i^0 \widetilde{\chi}_j^0$ pairs
$(i,j = 2,3,4)$ in the MSSM --- 9 basic types exist which
may be augmented by up to three stripes. 
At right, the region of the Dalitz-like plot can be broken up into 
the sections labeled $\alpha,\beta,...,\kappa$.  Envelope-types
A, B, and C are possible.  If the reduction from considering --inos
produced in pairs to considering only individual --ino production rates
is valid, only the patterns shown here are possible.
The density of points inside each region of the Dalitz-like
plot is shown to be uniform for simplicity only.
}}
\label{boxes}   
\end{figure}

The possible Dalitz topologies from --ino decays to
lepton pairs in (\ref{inopairprod}) are built from 3 possible boxes
and 3 possible wedges taken individually, and hence 63 basic
combinations ($\Sigma^6_{i=1} C(6,i)$) of boxes and wedges
when considered all together, though of these many are
topologically equivalent --- only the 9 topologically distinct
patterns shown in the left square of Figure \ref{boxes} are possible.
The patterns can be profitably categorised by the outer envelope exhibited
($A$, $B$, or $C$ as shown in the right square of Figure \ref{boxes}).
Additional internal structure can then further sub-divide members of each 
envelope-type.
To the extent that the reduction discussed in the preceding paragraph
is applicable for (\ref{glupairprod}), the envelope-type then depends on
the relative {\em individual} production times leptonic decay rates for 
$\widetilde{\chi}_{2,3,4}^0$ (call these $r_{2,3,4}$).
If $r_4$ is appreciable from {\em both} parent coloured sparticles, then
box $A$ is obtained (if $r_2$ and/or $r_3$ are also sizable, boxes and
wedges {\em inside} of the box $A$ envelope are also present);
with $r_2 > r_4 \gg r_3$, regions $\delta$, $\eta$, and $\kappa$ of the
Dalitz plot (see Figure \ref{boxes}) are down in population density by
${\sim}{r_4}^2$, and thus negligibly populated --- resulting in wedge
$B$;
finally with $r_2 \sim r_3 > r_4$ regions $\eta$ and $\kappa$ again are
negligibly populated yielding pattern $C$.
Higgs-mediated --ino pair production would move beyond
{\em individual} --ino production rates and probe the fundamental
--ino--ino$^{(\prime )}$-Higgs vertices, perhaps even more fertile
subject-matter {\it vis-\`a-vis} application of the Dalitz-like technique
$\!\!\!\!$ \cite{inPrep} (despite the lower maximal rates attainable).

Most previous LHC studies $\!\!\!\!$ \cite{BaerTataMachine} of 
multi-lepton signals from gluino/squark cascade decays have concentrated
on discovering evidence for SUSY, not upon extracting information about
the sparticle spectrum from observed leptons' momenta.  Earlier
attempts to look at the spectra of invariant
masses for lepton pairs resulting from gluino cascade decays are presented 
in $\!\!\!\!$ \cite{Paige,PaigeII,PaigeIII,DMR}.  This work only
examined one-dimensional invariant mass spectra where the electrons and
muons were not distinguished.  Further, the work was
restricted\footnote{\cite{BaerTatHigh} also tried to obtain information on 
the --ino mass spectrum from a similar invariant mass reconstruction of
tau-lepton pairs copiously produced in gluino decays at very high
$\tan\beta$.  Here mention is made of the heavier --inos in addition to
$\widetilde{\chi}_2^0$.} (unlike the discovery search just mentioned) to
the pair production of only the second lightest neutralino,
$\widetilde{\chi}_2^0 \widetilde{\chi}_2^0$.  Similar restrictions are
found in previous studies of Higgs-mediated --ino pair 
production $\!\!$ \cite{Higgs_to_Z2Z2_M,Higgs_to_Z2Z2_F}.
In fact, in $\!\!\!\!$ \cite{Higgs_to_Z2Z2_F} a Dalitz-like plot was
presented, but the $\widetilde{\chi}_2^0 \widetilde{\chi}_2^0$-only
condition meant that only a box was possible.  Thus the current work is
novel for its inclusion of the heavier --inos, $\widetilde{\chi}_{3}^0$
and $\widetilde{\chi}_{4}^0$, 
the presence of which leads to a far richer variety of possible decay 
topologies for study via the Dalitz-like method.
In addition, inclusion of the heaviest --ino states makes it more
comfortable to construct sparticle spectra with slepton masses near or
even below the heavier --inos.  Such a sparticle mass hierarchy can
greatly enhance the leptonic decay modes of the --inos\cite{BaerTata} 
--- leading to far larger signal event rates.


Decays of an --ino into a pair of same-flavour, oppositely-charged leptons
plus the LSP may proceed through either two- or three-body processes
with gauge boson or slepton intermediates; 
{\ie},
\begin{eqnarray}
 & \widetilde{\chi}_i^0 \to \{ Z^0,Z^{*0} \} + \widetilde{\chi}_1^0 
\to \ell^+\ell^- + \widetilde{\chi}_1^0 \\
\hbox{or} \;\;\; 
 & \widetilde{\chi}_i^0 
\to \ell^\pm + \{ \tilde{\ell}^\mp,\tilde{\ell}^{*\mp} \}  
\to \ell^+\ell^- + \widetilde{\chi}_1^0 \;\;\;\;\;\; ,
\end{eqnarray}
where the two-body decays occur through an on-mass-shell
$Z^0$-boson or slepton and the three-body decays occur when the 
$Z^0$-boson or slepton are off-mass-shell.
The $M(\ell^+\ell^-)$ spectra from the different decay processes
differ markedly.  If the decay is via an on-shell $Z^0$, then the 
lepton pair reconstructs the $Z^0$ and the spectrum is a sharp spike
at $M_Z$.  If the decay is via an on-shell slepton, then the 
$M(\ell^+\ell^-)$ spectrum is basically triangular
with a sharp rise in the number of event culminating at 
$\!\!\!\!$ \cite{Paige}
\begin{equation}
\label{two-body}
M(\ell^+\ell^-) < m_{\widetilde{\chi}_i^0} 
\sqrt{ 1 - \left(\frac{m_{\tilde{\ell}}}
                      {m_{\widetilde{\chi}_i^0}}\right)^{\!\!\! 2}}
\sqrt{ 1 - 
\left(\frac{m_{\widetilde{\chi}_1^0}}
           {m_{\tilde{\ell}}}\right)^{\!\!\! 2}} \;\;\; .
\end{equation}
Finally, if the decay is a three-body one via an off-shell $Z^0$
or slepton, then the $M(\ell^+\ell^-)$ spectrum is less sharply peaked
toward the high end, but extends up to 
\begin{equation}
M(\ell^+\ell^-) < 
m_{\widetilde{\chi}_i^0} - m_{\widetilde{\chi}_1^0} \;\;\; .
\end{equation}

Regardless of whether the heavy states decay 
through two- or three-body decays, the distribution of
dilepton invariant masses will be roughly
triangular ({\ie}, more decays occur toward the endpoint). This has two
immediate consequences: 1) hard kinematical edges in the Dalitz-like plot 
should be easy to identify since more of the event distribution is
pushed up against the endpoint;
2) the distribution of points inside the boxes and wedges will not be
uniform but can be fitted against an appropriate combination of
triangular distributions: hence ratios of different --ino 
production cross sections contributing to an observed topology may be
determined by comparing the number of points in different regions of the
Dalitz-like plot.  

\begin{figure}[t]
\dofig{3.00in}{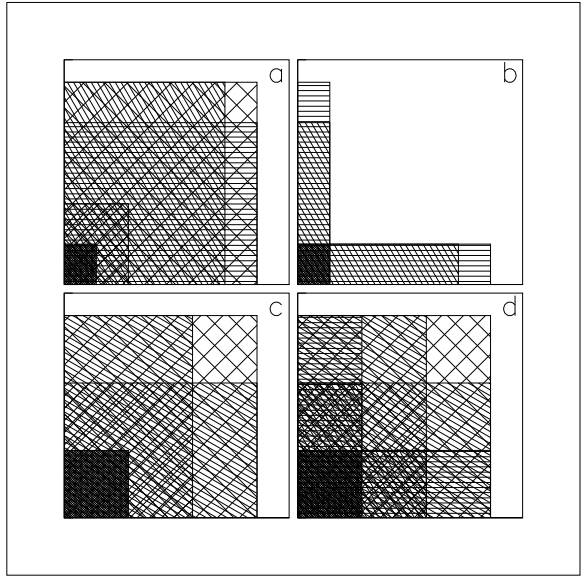}
\caption{{\small
Examples of how adding stripes affects the topologies shown in the left
square of Figure \ref{boxes}:
Some examples of one stripe added
a) one stripe added to basic type \#6,
b) one stripe added to \#2,
c) one stripe added to \#3,
and d) two stripes added to \#3 .
}}
\label{ex-boxes}
\end{figure}

On the other hand, one or both of the initially-produced heavy --inos
may not decay directly into the LSP plus leptons.  Cascading decay chains
including 
\begin{equation} 
\widetilde{\chi}_4^0 \to \widetilde{\chi}_3^0 +  \ell^+\ell^-
\;\; ,
\;\;\;\;
\widetilde{\chi}_4^0 \to \widetilde{\chi}_2^0 +  \ell^+\ell^-
\;\;\;\;
\&
\;\;\;\;
\widetilde{\chi}_3^0 \to \widetilde{\chi}_2^0 +  \ell^+\ell^-
\label{cascadeinodec}
\end{equation}
decays are also possible.   If present, these will add the afore-mentioned
stripes to the Dalitz-like plots.
Note that up to three stripes are possible
(these augment the 9 basic patterns shown in the left square of 
Figure \ref{boxes}).
Figure \ref{ex-boxes} illustrates how the 9 basic patterns of Figure
\ref{boxes} may be altered by the presence of stripes to further
enrich the number of possible Dalitz-like plot topologies. 
Note that among the topologies shown here, only in the case of Figure
\ref{ex-boxes}a can the observed hard edges in the Dalitz-like plot 
be unambiguously linked to specific --ino --ino pairs including all
the decaying --inos in the MSSM:
the three boxes must correspond to
$\widetilde{\chi}_2^0 \widetilde{\chi}_2^0$,
$\widetilde{\chi}_3^0 \widetilde{\chi}_3^0$, and
$\widetilde{\chi}_4^0 \widetilde{\chi}_4^0$ modes, while the stripe
must correspond to $\widetilde{\chi}_4^0$ cascading through one
of the other --inos (whether or not this is adequate to 
reconstruct all the mass differences in the 
complete --ino mass spectrum depends on the r\^oles played by the
sleptons).
One can imagine quite elaborate decay chains, with
$\widetilde{\chi}_4^0 \to \widetilde{\chi}_3^0 \to \widetilde{\chi}_2^0
\to \widetilde{\chi}_1^0$ for instance.  However, such elaborate chains
are very unlikely to emerge from any reasonable or even allowed choice
of MSSM input parameters.
Further, each step in such elaborate decay chains
either produces extra visible particles in the final state
or one must pay the price of the BR to neutrino-containing states.  The
latter tends to make the contribution from such channels insignificant,
while the former, in addition to also being suppressed by the additional
BRs, {\em may} also be cut (or enhanced) if extra restrictions are placed
on the final state composition in addition to demanding an $e^+e^-$ pair
and a $\mu^+\mu^-$ pair.
Another caveat is that decays with extra missing energy (carried off by
neutrinos, for example) or missed particles can further smear the endpoint.

To the discussion of leptonic --ino decays must be added the caveat
that there are other possible $\widetilde{\chi}_i^0$ decay modes where 
each lepton pair ($e^+e^-$ or $\mu^+\mu^-$) does not emerge from three-body 
or multiple two-body decays of a single --ino ---
exactly two leptons of different flavour can be obtained from the same
--ino (as noted in an earlier footnote).
For example, consider the following decay chain that includes a chargino
intermediate:
\begin{equation}
\widetilde{\chi}_i^0 \to \ell^+\nu +  \widetilde{\chi}_1^- 
\to \ell^+\nu {\ell^{\prime}}^-{\bar{\nu}}^{\prime} + \widetilde{\chi}_1^0
\;\;\; .
\end{equation}
It is also possible for all four leptons to come from one of the 
initial --inos while the other --ino yields no leptons.
This can occur, for example, if one --ino decays via
\begin{equation}
\widetilde{\chi}_i^0 \to \ell^+\ell^- \widetilde{\chi}_k^0
\to \ell^+\ell^-  \widetilde{\chi}_1^0 {\ell^{\prime}}^+{\ell^{\prime}}^-
\;\; ,
\end{equation}
while the other decays as
\begin{equation}
\widetilde{\chi}_j^0 \to \nu\bar{\nu} \widetilde{\chi}_1^0
\;\;\;
\hbox{or}
\;\;\;
\widetilde{\chi}_j^0 \to q\bar{q} \widetilde{\chi}_1^0
\;\;\; .
\end{equation}
Again though such channels are at least somewhat suppressed by the
additional required BRs.

Finally, note that 
if the decays proceed via a chain of two-body decays including an 
on-mass-shell slepton, then the edge positions of the topological 
shapes will depend on the mass of the slepton involved, as
seen\footnote{Note that in (\ref{two-body}) $m_{\tilde{\ell}}$  
is the physical slepton mass (or masses, if more than one intermediate
slepton is possible), not the soft slepton mass input,
$m_{\tilde{\ell}_R}$, defined in the following paragraph.  The physical
mass and the soft input may differ by several GeV or so.} in
(\ref{two-body}), and this may give rise to an asymmetry 
(noted for one-dimensional endpoints in $\!\!\!\!$ \cite{PaigeII})
in the Dalitz-like plot if the sleptons are not degenerate in
mass:  boxes will become rectangles and wedges will no longer be
symmetric under the exchange of axes.
Whether or not such deviations from boxes and wedges are discernible
depends on the slepton mass splittings.  The present work will not address 
this issue (degenerate or nearly-degenerate selectron and smuon masses
will be assumed).  Evidence for direct slepton pair production may either 
be useful in determining what is going in --ino--ino pair production 
processes if sleptons have such low masses as to be produced with
sufficient rates, or sleptons may be more massive and thus have low direct
pair production rates so that the --ino--ino event topologies and rates may 
shed light on the else-wise inaccessible slepton sector of the model.  

In this study, sleptons will be kept fairly light so as to enhance the
leptonic decay modes of the --inos $\!\!\!\!$ \cite{BaerTata}.  The
experimental limits from LEP on the slepton masses are\cite{W1LEP2} 
$m_{
{\tilde{   e}_1}
({\tilde{ \mu}_1})
[{\tilde{\tau}_1}]} \ge 99(91)[85]\, \hbox{GeV}$
and $m_{\tilde{\nu}} \ge 43.7\, \hbox{GeV}$.
To try to avoid producing leptons that are too soft, 
the charged sleptons (and the sneutrinos) 
are set sufficiently higher in mass than the LSP, which is the 
lightest neutralino, $\widetilde{\chi}_1^0$, as mentioned earlier.
Soft SUSY-breaking inputs are further simplified by assuming a
flavour-diagonal slepton sector with $m_{\tilde{\ell}_R} =
m_{\tilde{\ell}_L}$ and vanishing
trilinear `$A$-terms'.  This effectively reduces the slepton sector to one 
soft SUSY-breaking input mass (identified with 
$m_{\tilde{e}_{\scriptscriptstyle R}}
= m_{\tilde{\mu}_{\scriptscriptstyle R}}
\equiv m_{\tilde{\ell}_R}$)
in the analyses that follow.
These choices may be sub-optimal, especially since --ino decay modes to
sneutrinos (which only depend on $m_{\tilde{\ell}_L}$) tend to be
`spoiler' modes most often yielding only neutrinos in their decays and
no charged leptons, and nothing prevents choosing a more complex set of 
inputs for which a topological analysis may yield even more information.
Since --ino decays to tau-leptons are generally not anywhere near as
beneficial as are $\hbox{--ino}$ decays to electrons or muons,
it would be even better if the stau inputs were significantly above those
of the first two generations, thus the soft stau mass inputs are 
somewhat arbitrarily fixed to be $100\, \hbox{GeV}$ above the degenerate 
soft mass input chosen for the first two generations.  

Since the soft mass inputs for selectrons and smuons are degenerate, 
the masses of the actual physical sleptons will also be {\em nearly} 
degenerate.  With $m_{\tilde{\ell}_L} = m_{\tilde{\ell}_R}$
and $A_{\ell} = 0$, the physical slepton masses are 
given by
\begin{eqnarray}
\label{phys-sleps}
\!\!\!\!\!\!\!\! 
m^2_{\tilde{\ell}_{2,1}} & \!\!\! = \!\!\! & m^2_{\ell_R}
+ m^2_\ell - \frac{1}{4}M^2_Z\cos 2 \beta 
\pm \left[ \hbox{$\frac{1}{16}$} \left( 4M^2_W - 3M^2_Z \right)^2 
\cos^2 2 \beta 
+ m^2_\ell \mu^2\tan^2\beta  \right]^{\frac{1}{2}} \!\! , \\
\!\!\!\!\!\!\!\!
m^2_{\widetilde{\nu}} & \!\!\! = \!\!\! & m^2_{\ell_R} +
\frac{1}{2}M_Z^2\cos 2\beta \, .
\end{eqnarray}
The level at which the degeneracy is broken will be 
shown in some of the plots to follow; however, it remains too small
to be quantitatively analysed.  Thus the question of how non-degenerate
selectron and smuon masses can affect the observed topologies 
will not be probed in this first realistic simulation.

\section*{Detailed Study of Representative Points}

\begin{table}[t!]
     \caption{Relevant sparticle masses (in GeV) for Points 
   $A$, $B$ and $C$.}
     \begin{center}
     \begin{tabular}{|l||l|l|l|} \hline 
    &  A & B & C\\ \hline 
  ${\widetilde\chi}^0_1$  
           & \phantom{$1\!\!$} $93.9$ &  $186.4$ & $113.4$ \\ \hline 
  ${\widetilde\chi}^0_2$   
          & $136.7$ &  $248.8$ & $175.8$ \\ \hline
  ${\widetilde\chi}^0_3$   
           & $167.5$ & $257.8$ & $209.4$ \\ \hline
  ${\widetilde\chi}^0_4$   
          & $236.9$ &  $422.6$ &  $295.2$ \\ \hline
  ${\widetilde\chi}^\pm_1$ 
          & $136.4$ &  $238.3$ & $168.2$  \\ \hline
  ${\widetilde\chi}^\pm_2$ 
          & $238.2$ &  $422.6$ & $295.1$  \\ \hline\hline
  $m_{\widetilde{\nu}}$
          & $136.60$ & $189.38$ & $135.47$ \\ \hline
  $m_{\widetilde{e}_1}$
          & $155.82$ & $204.74$ & $156.28$  \\ \hline
  $m_{\widetilde{\mu}_1}$
          & $155.75$ & $203.76$ & $154.67$  \\ \hline
  $m_{\widetilde{e}_2}$
          & $156.71$ & $205.47$ & $157.24$  \\ \hline
  $m_{\widetilde{\mu}_2}$
          & $156.78$ & $206.44$ & $158.83$  \\ \hline
  $m_{\widetilde{e}_2} - m_{\widetilde{e}_1}$
         & \phantom{$1$} $0.89$ & \phantom{$1$} $0.73$ 
         & \phantom{$1$} $0.96$  \\ \hline
  $m_{\widetilde{\mu}_2} - m_{\widetilde{\mu}_1}$
         & \phantom{$1$} $1.03$ & \phantom{$1$} $2.68$ 
         & \phantom{$1$} $4.16$  \\ \hline
       \end{tabular}
    \end{center}
 \label{tab:masses}
\end{table}

To illustrate this technique, three points in the MSSM parameter space
with representative topologies were chosen for simulation
utilizing the Monte Carlo package $\!\!\!\!$
\cite{HERWIG} HERWIG 6.5.  With common inputs
of $M_{A^0} = 700\, \hbox{GeV}$, $m_{\tilde{g}} = 400\, \hbox{GeV}$,
and $m_{\tilde{q}} = 500\, \hbox{GeV}$ (for all soft squark mass inputs), 
the three points are
\newline
$\bullet$ Point $A$:
$\tan\beta = 5$,
$M_{2}(M_{1}) = 200(100)\, \hbox{GeV}$,
$\mu = -150\, \hbox{GeV}$,
$m_{\tilde{\ell}_{\scriptscriptstyle R}} = 150\, \hbox{GeV}$.
\newline
$\bullet$ Point $B$:
$\tan\beta = 20$,
$M_{2}(M_{1}) = 400(200)\, \hbox{GeV}$,
$\mu = -250\, \hbox{GeV}$,
$m_{\tilde{\ell}_{\scriptscriptstyle R}} = 200\, \hbox{GeV}$.
\newline
$\bullet$ Point $C$:
$\tan\beta = 30$,
$M_{2}(M_{1}) = 250(125)\, \hbox{GeV}$,
$\mu = 200\, \hbox{GeV}$,
$m_{\tilde{\ell}_{\scriptscriptstyle R}} = 150\, \hbox{GeV}$.
\newline
\noindent
Note that gaugino unification at a high (GUT) scale
is assumed for the EW gauginos, so $M_{1}$ is not independent of
$M_{2}$ ($M_{1} \simeq \frac{1}{2}M_{2}$ at the EW scale).
However, this restriction is relaxed for the gluino mass, which is taken
as independent of the EW gaugino masses.  The sparticle mass spectrum
for these three points is shown in Table \ref{tab:masses}. 
Masses do not include radiative corrections, which are generally small.
The shown physical slepton masses are also salient numbers to the Dalitz
plot analyses as they enter into (\ref{two-body}) and also largely
control the leptonic BRs of the --inos.  Note that if left-right    
sfermion mixing --- the $m^2_\ell \mu^2\tan^2\beta$ term in
Eqn.~(\ref{phys-sleps}) --- is neglected, the mass splitting of the
smuons becomes equal to that of the sleptons, and thus is evidently
sometimes markedly under-estimated.
Unfortunately, the physical slepton masses input (via ISASUSY 7.58
$\!\!\!\!\!$ \cite{ISAJET}) into the HERWIG simulations do neglect this
mixing.  This will be seen in the Dalitz-like plots shown later.

\begin{table}[t!]
  \vskip -0.5cm
     \caption{{\small
    Chargino and neutralino production and decay properties for 
    Points $A$, $B$ and $C$ obtained from ISAJET(ISASUSY) 7.58:  
    gluino \& squark BRs to neutralino/chargino and the
    inclusive BR for obtaining any number of leptons ($\ell=e,\mu$)
    from a neutralino/chargino (includes neither $\tau$-leptons
    nor leptons from $\tau$-decays).  Squark decays include decays 
    via a gluino; here $X$ denotes some number of quarks and gluons.
    Reverse the signs given for the charginos for anti-squark decays. 
    ``$-$'' means no BR at tree level.
    }}
     \begin{center}
     \begin{tabular}{|l||l|l|l|} \hline
   BRs for & Point $A$ & Point $B$ & Point $C$ \\ \hline \hline
   $\tilde{g} \rightarrow {\widetilde\chi}^0_1 \; g/q\bar{q}$
        & $.23$ &  $.61$ & $.34$ \\  
   $\widetilde{u}_L / \widetilde{d}_L /
       \widetilde{u}_R / \widetilde{d}_R
                       \rightarrow {\widetilde\chi}^0_1 X $
            & $.11$/$.13$/$.34$/$.26$ 
            & $.49$/$.54$/$.64$/$.62$ 
            & $.18$/$.22$/$.43$/$.36$  \\ 
   \hline \hline
   $\tilde{g} \rightarrow {\widetilde\chi}^0_2 \; g/q\bar{q}$
            & $.19$ &  $.18$ & $.25$ \\ 
   $\widetilde{u}_L / \widetilde{d}_L /
       \widetilde{u}_R / \widetilde{d}_R
                       \rightarrow {\widetilde\chi}^0_2 X $
            & $.15$/$.13$/$.19$/$.19$
            & $.18$/$.16$/$.18$/$.18$
            & $.21$/$.19$/$.24$/$.25$  \\
   \hline 
    $\;\;$ ${\widetilde\chi}^0_2
                \rightarrow 0\ell / 1\ell / 2\ell$
           & $\;\;$ $.76$/${\sim}10^{-7}\!$/$.25$  
          & $\;\;$ $.001$/$-$/$.999$ 
           & $\;\;$ $.305$/$-$/$.695$ \\
          \hline \hline
   $\tilde{g} \rightarrow {\widetilde\chi}^0_3 \; g/q\bar{q}$
               & $.038$ & $.11$ & $.10$ \\
   $\widetilde{u}_L / \widetilde{d}_L /
       \widetilde{u}_R / \widetilde{d}_R
                       \rightarrow {\widetilde\chi}^0_3 X $
            & $.02$/$.03$/$.03$/$.04$
            & $.09$/$.09$/$.09$/$.10$
            & $.05$/$.06$/$.08$/$.09$  \\
   \hline
    $\;\;$ ${\widetilde\chi}^0_3 
                  \rightarrow 0\ell/ 1\ell / 2\ell $
           & $\;\;$ $.93$/$.004$/$.07$  
           & $\;\;$ $.76$/$.00045$/$.24$
           & $\;\;$ $.80$/$.005$/$.19$
        \\
    \phantom{$\;\;$ ${\widetilde\chi}^0_3$ } 
           $/ 3\ell / 4\ell $
           & $\;\;$ /${\sim}10^{-9}$/$.002$
           & $\;\;$ /$-$/$-$
           & $\;\;$ /$-$/$3 \times 10^{-5}$
         \\ \hline \hline
   $\tilde{g} \rightarrow {\widetilde\chi}^0_4 \; g/q\bar{q}$
           & $.11$ & $\; 0 \;$ & $.024$ \\ 
   $\widetilde{u}_L / \widetilde{d}_L /
       \widetilde{u}_R / \widetilde{d}_R
                       \rightarrow {\widetilde\chi}^0_4 X $
            & $.15$/$.17$/$.09$/$.10$
            & $.03$/$.03$/
            $\!\!\! {\scriptstyle{ {\sim}10^{-5} / {\sim}10^{-5} }}$
            & $.09$/$.10$/$.02$/$.02$  \\
   \hline
    $\;\;$ ${\widetilde\chi}^0_4
                  \rightarrow 0\ell/ 1\ell / 2\ell $
           & $\;\;$ $.63$/$.094$/$.26$
           & $\;\;$ $.46$/$.16$/$.38$
           & $\;\;$ $.47$/$.20$/$.33$
    \\
    \phantom{$\;\;$ ${\widetilde\chi}^0_4$ \quad } 
                  $/ 3\ell / 4\ell  $
           & $\;\;$  /${\sim}10^{-8}\!$
                          /$.019$
           & $\;\;$     /$.0004$
                          /$.0013$
           & $\;\;$     /$.00008$
                          /$.0004$ 
    \\
    \phantom{$\;\;$ ${\widetilde\chi}^0_4$ \quad }
                  $ / 5\ell / 6\ell $
           & $\;\;$\quad
                          /$-$/$-$
           & $\;\;$ 
                          /${\sim}10^{-11}\!$
                          /${\sim}10^{-7}$
           & $\;\;$
                          /$-$
                          /${\sim}10^{-7}$
        \\ \hline \hline
   $\tilde{g} \rightarrow {\widetilde\chi}^\pm_1 \; q\bar{q}^{\prime}$
      & $.28$ &  $.11$ & $.26$ \\ 
   $\widetilde{u}_L / \widetilde{d}_L /
       \widetilde{u}_R / \widetilde{d}_R
                       \rightarrow {\widetilde\chi}^+_1 X $
            & $.25$/$.07$/$.11$/$.13$
            & $.12$/$.05$/$.05$/$.05$
            & $.25$/$.07$/$.10$/$.12$  \\
   $\widetilde{u}_L / \widetilde{d}_L /
       \widetilde{u}_R / \widetilde{d}_R
                       \rightarrow {\widetilde\chi}^-_1 X $
            & $.07$/$.12$/$.11$/$.13$
            & $.04$/$.07$/$.05$/$.05$
            & $.07$/$.16$/$.10$/$.12$  \\
   \hline 
   $\;\;$ ${\widetilde\chi}^\pm_1 
                   \rightarrow 0\ell / 1\ell$
           & $\;\;$ $.61$/$.39$  
           & $\;\;$ $.0023$/$.9977$
           & $\;\;$ $.0016$/$.9984$
     \\ \hline \hline   
   $\tilde{g} \rightarrow {\widetilde\chi}^\pm_2 \; q\bar{q}^{\prime}$
          & $.16$ &  $\; 0 \; $ & $.033$ \\ 
   $\widetilde{u}_L / \widetilde{d}_L /
       \widetilde{u}_R / \widetilde{d}_R
                       \rightarrow {\widetilde\chi}^+_2 X $
            & $.21$/$.04$/$.06$/$.07$
            & $.05$/$-$/$-$/$-$
            & $.14$/$.010$/$.013$/$.016$  \\
   $\widetilde{u}_L / \widetilde{d}_L /  
       \widetilde{u}_R / \widetilde{d}_R
                       \rightarrow {\widetilde\chi}^-_2 X $
            & $.04$/$.30$/$.06$/$.07$
            & $-$/$.07$/$-$/$-$
            & $.009$/$.20$/$.013$/$.016$  \\
   \hline 
   $\;\;$ ${\widetilde\chi}^\pm_2 
                  \rightarrow 0\ell/ 1\ell $
           & $\;\;$ $.11$/$.83$
           & $\;\;$ $.135$/$.756$
           & $\;\;$ $.060$/$.853$
     \\
   \phantom{ $\;\;$ ${\widetilde\chi}^\pm_2$ \quad } 
                  $ / 2\ell / 3\ell $
           & $\;\;$ /$.020$ /$.038$
           & $\;\;$ /$.084$/$.025$
           & $\;\;$ /$.065$/$.021$
           \\ 
   \phantom{ $\;\;$ ${\widetilde\chi}^\pm_2$ \quad }
                  $ / 4\ell / 5\ell $
           & $\;\;$  /${\sim}10^{-6}\!\!$
                          /${\sim}10^{-7}$
           & $\;\;$ /${\sim}10^{-9}\!\!$
                          /${\sim}10^{-6}$ 
           & $\;\;$ /${\sim} 10^{-7}\!\!$  
                          /${\sim}10^{-7}$ 
       \\ \hline 
       \end{tabular}
    \end{center}
 \label{tab:branchings}
\end{table}

Table \ref{tab:branchings} gives the BRs for squarks and gluinos to 
decay into charginos and neutralinos, and for charginos and neutralinos
to decay into final states with any number of leptons.  These BRs
were calculated using 
ISAJET(ISASUSY) 7.58 $\!\!\!\!$ \cite{ISAJET}\footnote{This is the version
incorporated into HERWIG 6.5; however, results sometimes differ
significantly from those obtained with later versions of ISASUSY.}.   
Naively, one might expect neutralinos (charginos) to only produce states
with an even (odd) number of charged leptons.
This is incorrect since combinations of quarks may also be produced in the 
decay chains, and said quark combinations can have a non-zero net charge.  
Note the significant BRs for $\widetilde{\chi}_2^{\pm} \rightarrow 2\ell$
in Table \ref{tab:branchings}.
As Figure 1(c) clearly shows, quarks are expected even before
neutralino/chargino decays are considered --- demanding hadronically-quiet
events is not an option in this case (but may be with the other production
modes), in fact just the opposite is most effective:
a minimum jet requirement will in fact be employed in the analysis to
follow.  
The neutralino and chargino BRs to $n\ell$ ($n=0,1,...$) final states
given in Table \ref{tab:branchings} include neither $\tau$-leptons nor
$\ell$s from $\tau$-decays.  Also, no demands are made on the leptonic
$p_T$ or $| \eta |$ values.

\begin{table}[h]
     \vskip -0.7cm
     \caption{Percentage contributions to $4\ell$ events
   from the various neutralino/chargino pair production modes 
   for MSSM Points $A$, $B$ and $C$.  
   Numbers in parentheses only consider gluino pair production. }
     \begin{center}
     \begin{tabular}{|l r@{}l r@{}l|lr@{}l r@{}l|lr@{}l r@{}l|}
\hline 
   \multicolumn{5}{|c}{Point $A$} &
   \multicolumn{5}{|c|}{Point $B$} &
   \multicolumn{5}{c|}{Point $C$} \\ \hline
  ${\widetilde\chi}^0_2{\widetilde\chi}^0_4$ 
$\,$ & $27.$&$45$\% & $\!\!\!$($25.$&$2$\%)
& ${\widetilde\chi}^0_2{\widetilde\chi}^0_2$ 
$\,$ & $73.$&$3$\% & $\!\!\!$($76.$&$3$\%)
& ${\widetilde\chi}^0_2{\widetilde\chi}^0_2$ 
$\,$ & $66.$&$6$\% & $\!\!\!$($72.$&$7$\%)
  \\
  ${\widetilde\chi}^0_2{\widetilde\chi}^0_2$ 
$\,$ & $14.$&$5$\% & $\!\!\!$($16.$&$7$\%)
& ${\widetilde\chi}^0_2{\widetilde\chi}^0_3$ 
$\,$ & $20.$&$3$\% & $\!\!\!$($22.$&$1$\%)
& ${\widetilde\chi}^0_2{\widetilde\chi}^0_3$ 
$\,$ & $13.$&$3$\% & $\!\!\!$($15.$&$9$\%)
  \\
  ${\widetilde\chi}^0_4{\widetilde\chi}^0_4$ 
$\,$ & $11.$&$45$\% & $\!\!\!$($8.$&$6$\%)
& ${\widetilde\chi}^0_2{\widetilde\chi}^0_4$ 
$\,$ & $3.$&$2$\%  & $\!\!\!$($0.$&$0$\%)
& ${\widetilde\chi}^0_2{\widetilde\chi}^0_4$ 
$\,$ & $12.$&$5$\% & $\!\!\!$($6.$&$6$\%)
  \\
  ${\widetilde\chi}^0_1{\widetilde\chi}^0_4$ 
$\,$ & $7.$&$8$\% & $\!\!\!$($7.$&$6$\%)
& ${\widetilde\chi}^0_3{\widetilde\chi}^0_3$ 
$\,$ & $1.$&$4$\% & $\!\!\!$($1.$&$6$\%)
& ${\widetilde\chi}^{\pm}_2{\widetilde\chi}^0_2$ 
$\,$ & $2.$&$7$\% & $\!\!\!$($1.$&$8$\%)
  \\
  ${\widetilde\chi}^+_2{\widetilde\chi}^-_2$ 
$\,$ & $7.$&$4$\% & $\!\!\!$($6.$&$25$\%)
& ${\widetilde\chi}^{\pm}_2{\widetilde\chi}^0_2$ 
$\,$ & $0.$&$9$\% & $\!\!\!$($0.$&$0$\%)
& ${\widetilde\chi}^0_3{\widetilde\chi}^0_4$ 
$\,$ & $1.$&$3$\% & $\!\!\!$($0.$&$7$\%)
  \\
  ${\widetilde\chi}^{\pm}_1{\widetilde\chi}^0_4$ 
$\,$ & $6.$&$6$\% & $\!\!\!$($8.$&$7$\%)
& ${\widetilde\chi}^0_3{\widetilde\chi}^0_4$ 
$\,$ & $0.$&$5$\% & $\!\!\!$($0.$&$0$\%)
& ${\widetilde\chi}^{\pm}_1{\widetilde\chi}^{\mp}_2$
$\,$ & $0.$&$74$\% & $\!\!\!$($0.$&$9$\%) 
  \\
  ${\widetilde\chi}^{\pm}_2{\widetilde\chi}^{\pm}_2$ 
$\,$ & $6.$&$5$\% & $\!\!\!$($6.$&$25$\%) 
& ${\widetilde\chi}^{\pm}_2{\widetilde\chi}^0_3$ 
$\,$ & $0.$&$1$\% & $\!\!\!$($0.$&$0$\%)
& ${\widetilde\chi}^0_3{\widetilde\chi}^0_3$
$\,$ & $0.$&$68$\% & $\!\!\!$($0.$&$9$\%)
  \\
  ${\widetilde\chi}^{\pm}_1{\widetilde\chi}^{\mp}_2$ 
$\,$ & $4.$&$4$\% & $\!\!\!$($5.$&$15$\%)
& ${\widetilde\chi}^{\pm}_1{\widetilde\chi}^{\mp}_2$ 
$\,$ & $0.$&$1$\% & $\!\!\!$($0.$&$0$\%)
& ${\widetilde\chi}^{\pm}_1{\widetilde\chi}^{\pm}_2$ 
$\,$ & $0.$&$67$\% & $\!\!\!$($0.$&$9$\%)
  \\
  ${\widetilde\chi}^{\pm}_1{\widetilde\chi}^{\pm}_2$ 
$\,$ & $4.$&$3$\% & $\!\!\!$($5.$&$15$\%)
& ${\widetilde\chi}^{\pm}_1{\widetilde\chi}^{\pm}_2$ 
$\,$ & $0.$&$08$\% & $\!\!\!$($0.$&$0$\%)
& ${\widetilde\chi}^0_4{\widetilde\chi}^0_4$ 
$\,$ & $0.$&$5$\% & $\!\!\!$($0.$&$15$\%)
  \\
  ${\widetilde\chi}^{\pm}_2{\widetilde\chi}^0_4$ 
$\,$ & $2.$&$9$\% & $\!\!\!$($2.$&$9$\%)
   &  & & & &
& ${\widetilde\chi}^{\pm}_2{\widetilde\chi}^0_3$ 
$\,$ & $0.$&$3$\% & $\!\!\!$($0.$&$2$\%)
  \\
  ${\widetilde\chi}^0_3{\widetilde\chi}^0_4$ 
$\,$ & $2.$&$6$\% & $\!\!\!$($2.$&$4$\%)
   &  & & & &
& ${\widetilde\chi}^{\pm}_2{\widetilde\chi}^0_4$ 
$\,$ & $0.$&$3$\% & $\!\!\!$($0.$&$1$\%)
  \\
  ${\widetilde\chi}^{\pm}_2{\widetilde\chi}^0_2$ 
$\,$ & $1.$&$9$\% & $\!\!\!$($2.$&$3$\%)
   & & & & &
&  ${\widetilde\chi}^+_2{\widetilde\chi}^-_2$ 
$\,$ & $0.$&$24$\% & $\!\!\!$($0.$&$05$\%)
  \\
  ${\widetilde\chi}^0_2{\widetilde\chi}^0_3$ 
$\,$ & $1.$&$8$\% & $\!\!\!$($2.$&$1$\%)
   & & & & & 
&  ${\widetilde\chi}^{\pm}_2{\widetilde\chi}^{\pm}_2$ 
$\,$ & $0.$&$15$\% & $\!\!\!$($0.$&$05$\%)
  \\
  ${\widetilde\chi}^0_1{\widetilde\chi}^0_3$ 
$\,$ & $0.$&$2$\% & $\!\!\!$($0.$&$3$\%)
   & & & & &
   & & & & &
  \\
  ${\widetilde\chi}^{\pm}_1{\widetilde\chi}^0_3$ 
$\,$ & $0.$&$1$\% & $\!\!\!$($0.$&$2$\%)
   & & & & &
   & & & & &
  \\
  ${\widetilde\chi}^{\pm}_2{\widetilde\chi}^0_3$ 
$\,$ & $0.$&$1$\% & $\!\!\!$($0.$&$2$\%)
   & & & & &
   & & & & &
  \\
  \hline
       \end{tabular}
    \end{center}
 \label{tab:percents}
\end{table}

Combining the leptonic BRs of the assorted neutralinos and charginos with
their production rates from decays of gluinos and the different squarks
yields Table \ref{tab:percents}.  The neutralino, chargino, or 
neutralino/chargino pair listed represents the first EW sparticles
produced in decays of the colored sparticles.  The EW sparticles can then
themselves decay into other EW sparticles.  `Mixed' production modes
($\tilde{g}\chi_i^0$, $\tilde{q}\chi_i^0$,
$\tilde{g}\chi_i^{\pm}$, $\tilde{q}\chi_i^{\pm}$)
are also included.  These mixed modes account for only about 
($2.7$\%, $1.0$\%, $1.4$\%) of the events for MSSM Point ($A$, $B$, $C$).  
Since there is less jet activity with the mixed modes, they are more 
likely to fail the minimum jet requirement.  HERWIG lacks the facilities
for giving the cross-sections for each separate $\tilde{g}\tilde{q}$,
$\tilde{q}\tilde{q}^{(\prime )}$, $\tilde{g}\chi$ and $\tilde{q}\chi$
process, so these were calculated using ISAJET 7.67 $\!\!\!\!\!$
\cite{ISAJET} with CTEQ5 $\!\!\!\!\!$ \cite{CTEQ6} parton distributions.  
Some tinkering with the HERWIG code was however able to yield values for
$\sigma(\sum \tilde{g}\tilde{q})$ and 
$\sigma(\sum \tilde{q}\tilde{q}^{(\prime )})$ 
as well as $\sigma(\tilde{g}\tilde{g})$.  ISAJET+CTEQ5 cross-sections
were virtually always found to be lower than those from 
HERWIG+CTEQ6.  The $\sigma(\sum \tilde{q}\tilde{q}^{(\prime )})$s
agreed to ${\sim}$5\% at the three MSSM parameter points, while 
the ISAJET+CTEQ5 $\sigma(\sum \tilde{g}\tilde{g})$s 
($\sigma(\sum \tilde{g}\tilde{q})$s)
were lower by roughly 10-20\% (5-10\%).  
Given that HERWIG and ISAJET differ in the scales adopted for the
parton distribution functions (which are also different here)
and for the evolution of coupling constants, the differences seen in  
these cross-sections are in fact quite modest.  Thus using ISAJET
rather than HERWIG values should not markedly effect the estimates
obtained.
For an integrated luminosity of $30\, \hbox{fb}^{-1}$
(equivalent to two or three years of low-luminosity performance 
at the LHC) ISAJET+CTEQ5 predicts approximately 60,000, 200,000 and
197,000 $4\ell$ events before any cuts are applied for MSSM Points
$A$, $B$ and $C$, respectively.  By contrast there are only
744, 471 and 750 $4\ell$ events from `direct' production of
charginos/neutralinos at the three points in parameter space
($1.2$\%, $0.2$\% and $0.4$\% of the coloured-sparticle cascade rates).

The percentages given in parentheses in Table \ref{tab:percents}
are when only production via gluinos is considered.  Here the reduction
in the possible topologies mentioned earlier\footnote{ The so-called
`second point warranting attention' in the section entitled
``Application to Sparticle Decays.''} applies.  For instance, 
look at the
${\widetilde\chi}^0_2{\widetilde\chi}^0_2$,
${\widetilde\chi}^0_3{\widetilde\chi}^0_3$ and
${\widetilde\chi}^0_2{\widetilde\chi}^0_3$ fractions (in parentheses) 
for Points $B$ and $C$ --- or the
${\widetilde\chi}^0_2{\widetilde\chi}^0_2$,
${\widetilde\chi}^0_4{\widetilde\chi}^0_4$ and
${\widetilde\chi}^0_2{\widetilde\chi}^0_4$ fractions for Point $A$ ---
labeling these as $r^2_{i}$, $r^2_{j}$ and $r_{ij}$, respectively
(they are proportional to the production cross-section times the $4$
lepton BR for the given --ino pair), we find that $r_{ij} =
2r_{i}r_{j}$.
Explicitly (ignoring the insignificant `mixed' production channels),
\begin{eqnarray}
r_{ij} = \hbox{BR}(2 \, \hbox{coloured sparticles} \rightarrow
\hbox{neutralino}_i \, + \, \hbox{neutralino}_j)
\nonumber \\
\;\;\;\;\;\;\;\;\;\;\;\;\;        
*  \; \hbox{BR}( \hbox{neutralino}_i + \hbox{neutralino}_j
\rightarrow 4 \, \hbox{leptons} ),
\end{eqnarray}
\vskip -0.4cm
while
\vskip -0.9cm
\begin{eqnarray}
r_i = \hbox{BR}( \hbox{gluino or squark} \rightarrow
\hbox{neutralino}_i) 
\;\;\;\;\;\;\;\;\;\;\;\;\;\;\;\;\;\;\;\;\;\;\;\;\;\;\;\;\;
\nonumber \\ 
\;\;\;\;\;\;\;\;\;\;\;\;\;  
* \; \hbox{BR}(\hbox{neutralino}_i \rightarrow 
\hbox{designated number of leptons}).
\end{eqnarray}
Under the assumptions that 
\newline
$\bullet$
1.  Each --ino came from a gluino;
\newline
$\bullet$ 
2.  Each --ino produced two leptons 
(presumably of the same flavour); 
\newline
then $r_{ij}$ factorises as 
\begin{eqnarray}
r_{ij} &=& \hbox{BR}(2 \, \hbox{gluinos} \rightarrow
\hbox{neutralino}_i \, + \, \hbox{neutralino}_j)
\;\;\;\;\;\;\;\;\;\;\;\;\;\;\;\;\;\;\;\;\;\;\;\;\;\;\;\;\;\;\;\;
\nonumber \\
& & \;\;\;\; *  \; \hbox{BR}( \hbox{neutralino}_i +
\hbox{neutralino}_j
\rightarrow 4 \, \hbox{leptons} 
\nonumber \\
\;\;\;\;\;\; &=& 2 \; * \;
\hbox{BR}(\hbox{gluino} \rightarrow \hbox{neutralino}_i) 
 \;  * \; \hbox{BR}(\hbox{neutralino}_i \rightarrow
       2\, \hbox{leptons})
\nonumber \\
& & \;\;\;\;\; * \;
\hbox{BR} ( \hbox{gluino} \rightarrow \hbox{neutralino}_j) 
\;  * \; \hbox{BR}(\hbox{neutralino}_j 
\rightarrow 2\, \hbox{leptons})
\nonumber \\
\; &=& 2 \, r_i \, r_j \;\; .
\end{eqnarray}
Checking the BRs for the various --ino pairs
from when one --ino produces $m$ leptons and the other produces $n$
leptons, where $m+n=4$, shows that only the $m=n=2$ case contributes
significantly.  The difference between the percentage when all
production modes are included and the percentage in parentheses thus
quantifies the deviation due to the squark production modes.  That these
two values generally do not differ by too much indicates that the
relationships among the --ino pair production rates expected
from gluino-only production do to a significant extent remain intact 
when squarks are included.  There is a caveat to this though:  here only
inclusive $4\ell$ events are tabulated with no cuts; squark events may 
contain more jets and thus a higher percentage of them may pass a minimum
jet number requirement. 

Consider a numerical example (to be compared later to results
extracted from simulations via the Dalitz-like technique):
for Point $C$, Table \ref{tab:percents} gives
\begin{equation}
 r_{23} : r_{24} : r_{34} = .133 : .125 : .013 = 10.2 : 9.6 : 1
\end{equation}
(or $r_{23} : r_{24} : r_{34} = .159 : .0660 : .00721
= 22.0 : 9.16 : 1 \; $
if only gluino pair production is considered).
{\em Assuming} the formula $r_{ij} = 2 r_i r_j$ holds, it follows that
$ r_2 : r_3 : r_4 = 19.00 : 1.04 : 1 $
 ($ r_2 : r_3 : r_4 = 22.00 : 2.404  : 1 $ with only gluino pair
production\footnote{
 The ratio of $\tilde{g}\tilde{g}$-production to
                 $\tilde{g}\tilde{q}$-
             and $\tilde{q}\tilde{q}^{(\prime )}$-production
      is crucial here, and this ratio is larger for HERWIG+CTEQ6
      than for ISAJET+CTEQ5.  So the latter would yield larger deviations 
      from $r_{ij} = 2r_ir_j$.  For such deviations to be taken as evidence 
      for squark-initiated processes, it needs to be shown that the measured 
      deviations exceed the uncertainties due to structure functions and 
      simulator cross-section estimates.}
--- this result matches the values obtainable from 
Table \ref{tab:branchings}).
Alternatively, the identical --ino pair values can be used, assuming
$r_{ii} = r_i^2$.   For Point $C$, Table \ref{tab:percents} then gives
\begin{equation}
 r_{22} : r_{33} : r_{44} = .666 : .0068 : .005 = 131.5 : 1.3 : 1
\end{equation}
(or $r_{22} : r_{33} : r_{44} = .727 : .009 : .0015
= 485.3 : 5.78 : 1 \; $
if only gluino pair production is considered).
These values yield 
$ r_2 : r_3 : r_4 = 11.47 : 1.16 : 1 $
 ($ r_2 : r_3 : r_4 = 22.03 : 2.404  : 1 $ with only gluino pair
production).  Note that the results considering only gluino pair
production agree, while the full results do not.  Thus disagreement
in such calculations indicates significant contributions from squark
production.

Roughly a third of the $4\ell$ events for Point $A$ come from 
production modes including charginos.  However, a substantial fraction of
these events will not have leptons in same-flavour, opposite-sign
pairs.  So their effect on this analysis will be diminished\footnote{In
fact, same-flavour, {\em like-sign} $4\ell$ events
({\it i.e.}, $e^{\pm}e^{\pm}{\mu}^{\mp}{\mu}^{\mp}$ events) could be used
to estimate the chargino contribution and then remove it.  This is
seen, though with a different rationale, in $\!\!\!\!$ 
\cite{PaigeII,PaigeIII}.}.
This does expose a minor weakness of the framework developed herein which
is built only for the neutralinos.  The chargino production contributions
for Points $B$ and $C$ are much smaller (${\sim}1$\% and ${\sim}5$\%,
respectively).  

Table \ref{tab:percents} is only expected to serve as a guideline
against which simulation results may be examined.
While this will prove useful in confirming the interpretations of
features on the Dalitz-like plots, it should be emphasised that this is  
information that the real experiments will not be able to access;
{\it i.e.}, with a simulation, we are of course able to choose what point
in the parameter space to simulate.

\section*{Numerical Simulations}

Events for 
$pp \to \tilde{g}\tilde{g},\tilde{g}\tilde{q},\tilde{q}\tilde{q},
\tilde{g}\widetilde{\chi},\tilde{q}\widetilde{\chi}$
were generated at the specific points in the MSSM parameter space 
just discussed using HERWIG 6.5 coupled with a detector
simulation which assumes a typical LHC experiment, as provided by 
private programs checked against results in the literature.  
The CTEQ6M $\!\!\!\!$ \cite{CTEQ6} set of structure functions was used in
conjunction with HERWIG 6.5 to determine the cross-sections.

In event selection and subsequent cuts, stress is put on keeping the cuts
reasonably general so that they will hopefully be applicable across a
large swath of the allowable parameter space.  These cuts can quite
probably be further honed once the first evidence(hints) 
of possible MSSM events is discerned, and the rather minimal selection
criteria used here certainly do not represent the optimal choice 
for the few points examined in detail in this work.
The actual criteria used are as follows: 
\begin{itemize}

\item $4\ell$ events: events are selected which have exactly four 
isolated leptons 
\newline
($\ell=e^{\pm}$ or $\mu^{\pm}$) 
with $|\eta^\ell|<2.4$ and $E_T^\ell >7,4\, \hbox{GeV}$ 
for $e^{\pm},\mu^{\pm}$, respectively.
The isolation criterion demands there be no tracks
(of charged particles) with $p_T > 1.5\, \hbox{GeV}$ in a cone of 
$r = 0.3\, \hbox{radians}$ around a specific lepton, and also that
the energy deposited in the electromagnetic calorimeter be less than
$3\, \hbox{GeV}$
for $0.05\, \hbox{radians} < r < 0.3\, \hbox{radians}$.

\item The four leptons must consist of exactly one
$e^+e^-$ pair and one $\mu^+\mu^-$ pair.

\item A cut on missing transverse energy: 
events must have $\slashchar{E}_T \ge 20\, \hbox{GeV}$.

\item Three or more jets must be present. 
Jets are defined by a cone algorithm with $r=0.4$
and must have $|\eta^j|<2.4$ and $E_T^j >20\, \hbox{GeV}$.

\end{itemize}
In addition to requiring the advertised lepton
make-up of the final state demanded by the Dalitz-like technique,
we apply the well-known 
canonical missing energy cut to select for SUSY events.  Further,
we require some minimum number of jets as noted.  Since production
mechanism (\ref{glupairprod}) typically generates considerable
jet activity, while the $H,A$ `background' does not,
such a cut is found to be particularly helpful.
There are no cuts on the momenta properties of the 
leptons (aside from demanding that they be hard and central enough).
This is consistent both with 
(a) the wish to paint all the leptons from the
signal events onto our Dalitz-esque canvass to show the richness of the
possible topologies, and with (b) the desire not to narrow the
applicability down to only a minor portion of the MSSM parameter space
currently experimentally viable that can satisfy such additional
restrictions.  The lepton isolation criterion is essential though to 
virtually eliminate enormous QCD backgrounds from events with 
leptonically-decaying $b$-quarks (such as from $t\bar{t}$ production).
Note also that gluino cascade decays are often rich in $b$-quarks
(particularly for higher values of $\tan\beta$); thus addition of a 
$b$-tagging requirement {\em might} further enhance the signal/background
ratio --- but probably at the loss of some significant fraction of the
signal events.  For present purposes such a cut was found unnecessary.
That, as will be shown next, the signal stands out over the backgrounds
with so few cuts attests to the robustness of this signature and to the 
potential to obtain Dalitz-like plots using realistic simulated data that
reflect the theoretical expectations discussed in the previous sections.

\begin{table}[t!]
     \vskip -0.7cm
\newcommand{\Lower}[1]{\smash{\lower 1.45ex \hbox{#1}}}
    \caption{{\small
     Number of events after the successive cuts defined in the
     text for MSSM Parameter Points $A$, $B$, and $C$ 
     (for $30\, \hbox{fb}^{-1}$).
     }}
    \begin{center}
    \begin{tabular}{|c|l||c|c|c|c|} \hline
    &{Process}
    & {$4\ell$ events}
    & { $e^+e^-$-$\mu^+\mu^-$ pairs }
    & {$E_T^{\rm{miss}}$} & $N_{jets} \ge 3$
\\
 \hline
 &  $Z^0Z^0$
        & 365 & 175   & 11 & 0    \\
    & $Z^0~$+ jet
      & 0   &  0   & 0    & 0  \\
   SM processes  &  $\bar{t}~H^+$, $t~H^-$
     & 1   &  1   & 1    & 0 \\
  (common) & $t\bar{t}$
	 & 0   &  0   & 0   & 0  \\
  &  $t\bar{t}Z^0$
        & 47  & 7   & 6   & 2 \\
  & $t\bar{t}h^0$
        & 4   & 1     & 1   & 0 \\ 
  &  total SM bkg.  
        & 417    &  184       &   19 & 2 \\ \hline  \hline
  &  $\widetilde{\chi}\, \widetilde{\chi}$ (direct)
      & 26  &  6     & 6    & 0    \\
 Point $A$  &  $\tilde{\ell}$,$\widetilde{\nu}$
        & 50  & 23   & 21    &  1  \\
  &   $A^0, H^0$
        & 29    &  5       &   5  & 0    \\ 
  &  $\tilde{g},\tilde{q}$ signal
        & 7628 & 2350 & 2292   & 2110 \\ \hline
\hline  
    &  $\widetilde{\chi}\, \widetilde{\chi}$ (direct)
       & 95  &  36     &  34  &   2   \\ 
 Point $B$  &  $\tilde{\ell}$,$\widetilde{\nu}$
        & 5  & 0   & 0    & 0   \\ 
  &  $A^0, H^0$
        & 85    &   26     &    26 & 4  \\ 
  &  $\tilde{g},\tilde{q}$ signal
        & 15652 & 7114 & 6883  & 5979  \\ \hline\hline
  &  $\widetilde{\chi}\, \widetilde{\chi}$ (direct)
      & 78  &  20     &  18   & 1     \\
 Point $C$ &  $\tilde{\ell}$,$\widetilde{\nu}$
        & 12  & 3   & 3  &  0    \\ 
  &   $A^0, H^0$
        & 292    &   114      &    109 & 5   \\
  &  $\tilde{g},\tilde{q}$ signal
        & 19897 & 8595 & 8357  & 7615  \\ \hline\hline
    \end{tabular}
    \end{center}
\label{tab:cuts}
\end{table}

Both $\tilde{g},\tilde{q}$ signal events and SM \& MSSM background events
were simulated at each point again assuming an integrated luminosity of
$30\, \hbox{fb}^{-1}$ to see how the expected features show up for a
realistic sample size.  The number of events passing each of the cuts
above for points $A$, $B$ and $C$ are listed in Table \ref{tab:cuts},
which clearly illustrates how effective even this limited set of cuts 
is at eliminating the SM backgrounds\footnote{SM background
events from $Z^0Z^0$ production would be concentrated around $M_Z$ were
they not eliminated by the three jet minimum requirement.  When the
other production mechanisms are considered in later works, such
a requirement will probably neither be possible nor desirable.
Then, though the relative number of SM background events passing cuts
may yet be quite low relative to the total number of signal events
in the plot, they can still lead to uncertainty in precisely locating
edges (particularly indistinct ones) that happen to be in the close
vicinity of $M_Z$.}.  
And, at least for the particular cases studied here in detail, a
sufficient percentage of signal events pass the cuts,
while the number of surviving events from SUSY ``background''
processes --- primarily
$pp\to \tilde{\chi}_i^{0,\pm} \tilde{\chi}_i^{0,\pm},
\tilde{\chi}_i^{0,\pm} \tilde{\chi}_j^{0,\pm}$ and $pp\to A^0,H^0$
--- is negligible.

Note that the `$4\ell$' rates given in the first column of
Table \ref{tab:cuts} are roughly an order of magnitude smaller
than the inclusive $4\ell$ rates predicted in the previous
section from the ISASUSY inputs.  Further investigation indicated that 
somewhat less than half of the events were lost when the lowest $E_T$ 
lepton failed the imposed minimum $E_T$ cut.  Other factors in the event 
(such as heavy quark decays) could also have yielded extra leptons, so that   
$4$-lepton events became $5$-lepton events; however, the number of 
$n>4$ lepton events was checked to be quite small.  Some events certainly
had leptons too close to the beam pipe, but, again, this is not expected
to be a major factor.  We are thus led to conclude that the majority of
the $4\ell$ events were removed due to the isolation requirements. 
The fact that, as we shall see, the simulation results, qualitatively
at least, track the values given in Table \ref{tab:percents} fairly well 
is consistent with this hypothesis (if the main factor had been the 
minimum $E_T$ cut, for instance, $\widetilde{\chi}^0_i\widetilde{\chi}^0_j$ 
events, where $i$ and/or $j$ is $2$, might have been highly preferentially 
eliminated).
Nonetheless, the large fraction of events removed and the subsequent
cuts applied caution against expecting a high degree of quantitative 
agreement between the simulation results and those of 
Table \ref{tab:percents} (as already noted).

\begin{figure}[t]
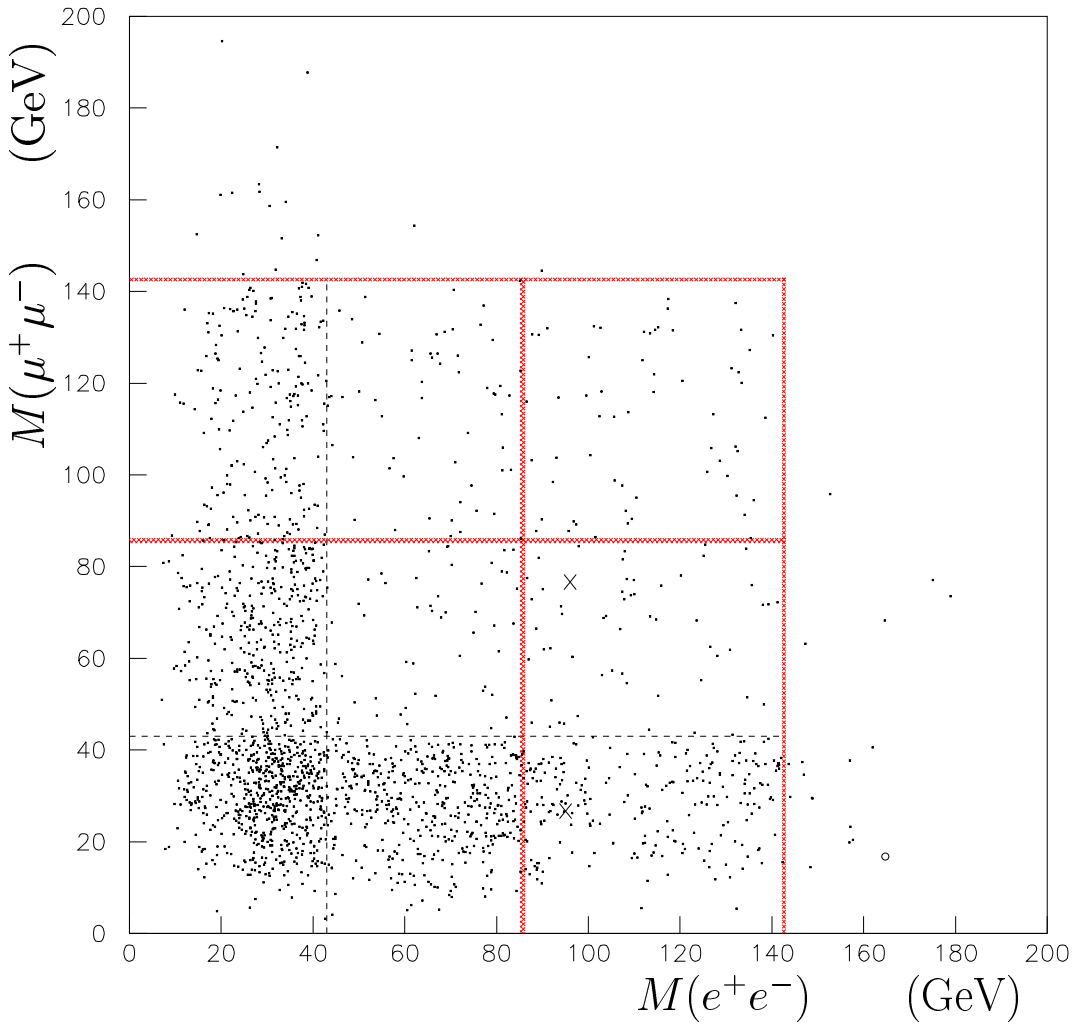

\vskip 3.1cm
\hspace{0.5cm}\dofig{14.0cm}{figure4.eps}
\vskip -9.25cm
\caption{{\small
$M(e^+e^-)$ {\it vs.} $\! M(\mu^+ \mu^-)$
Dalitz-like plot for MSSM Point $A$ assuming an integrated luminosity
of $30\, \hbox{fb}^{-1}$.
Signal events from gluino/squark production are denoted by dots,
SM background events by crosses and events from other SUSY processes
by open circles.
Shaded bands (dashed lines) indicate kinematical endpoints expected 
from two-body decays (three-body decays) based on ISASUSY.
}}
\label{pointA}
\end{figure}

Figure \ref{pointA} presents the Dalitz-like plot for 
MSSM parameter point $A$.
A `wedge inside of a box' topological structure is apparent 
(as per pattern $A$ in the right-side square of Figure \ref{boxes}), 
a clear indication that two pairs of --inos, 
$\widetilde{\chi}_i^0\widetilde{\chi}_j^0$ ($i < j$)
and $\widetilde{\chi}_j^0\widetilde{\chi}_j^0$,
are being produced at significant rates.
The (\ref{glupairprod}) production mechanism then 
demands that a $\widetilde{\chi}_i^0\widetilde{\chi}_i^0$
box also be present, the position of which overlaps with 
that of the low $M(e^+e^-)$, low $M(\mu^+ \mu^-)$ corner
of the wedge (as noted earlier, adding such a box is not viewed
as being topologically distinct).
A hard kinematical edge ({\ie}, the line in the
plot across which the density of points changes very rapidly)
at ${\sim}40$-$45\,\hbox{GeV}$ is very apparent.
The outer box seems to end at ${\sim}140\, \hbox{GeV}$ though there
are a small number of straggling points beyond this mostly at
high $M(e^+ e^-)$, low $M(\mu^+ \mu^-)$ and at
low $M(e^+ e^-)$, high $M(\mu^+ \mu^-)$.  Also
discernible inside the wedge are somewhat indistinct drops in 
population densities along both axes at ${\sim}85\,\hbox{GeV}$.

The shaded bands and dashed lines included in the plot show the
expected locations of hard edges based on the --ino and slepton mass
spectrum obtained from ISASUSY for Point $A$.  
The ${\sim}40$-$45\,\hbox{GeV}$ hard edge corresponds 
to the $42.8\, \hbox{GeV}$ $\widetilde{\chi}_2^0$-$\widetilde{\chi}_1^0$ mass
difference.  Here $\widetilde{\chi}_2^0$ is decaying through an
off-shell $Z^0$ or slepton, with 
$BR(\widetilde{\chi}_2^0 \rightarrow 
\widetilde{\chi}_1^0 \ell^+ \ell^-) = 0.245$
(very unlike the leptonic BR for the $Z^0$)
indicating that the off-shell sleptons are playing large r\^oles.
The other --inos decay mainly through on-mass-shell sleptons.
The outer edge at ${\sim}140\, \hbox{GeV}$ agrees with the endpoint for 
the two-body decay chain
$\widetilde{\chi}_4^0 \to \tilde{\ell}\ell \to \widetilde{\chi}_1^0 +
\ell^+\ell^-$ (though this is the actual decay channel for this
sparticle spectrum, in fact the two-body decay and three-body decay
endpoints differ by less than $1\, \hbox{GeV}$ in this case).   
So the outer box is from 
$\widetilde{\chi}_4^0\widetilde{\chi}_4^0$
production and the wedge is from
$\widetilde{\chi}_2^0\widetilde{\chi}_4^0$
(including an inner box from 
$\widetilde{\chi}_2^0\widetilde{\chi}_2^0$
production).  

The population changes at ${\sim}85\,\hbox{GeV}$ inside the wedge might
be interpreted as evidence for significant 
$\widetilde{\chi}_2^0 \widetilde{\chi}_3^0$ production, or as a `stripe'.  
In fact, they are due to the latter, and are associated with the decay
chain $\widetilde{\chi}_4^0 \to \tilde{\ell}\ell \to
\widetilde{\chi}_2^0 + \ell^+\ell^-$ which happens $22.8$\% of the time.  
The $\widetilde{\chi}_4^0$ decay
chain mentioned in the last paragraph occurs $65.8$\% of the time, 
and the remaining $10.6$\% of the $\widetilde{\chi}_4^0$ decays are
through $\widetilde{\chi}_1^{\pm}$.  Note that at least some 
{\it a priori} knowledge of the --ino mass spectrum and decay modes is 
required to designate this feature a stripe, showing
that such Dalitz-like plots do not always uniquely identify
the underlying --ino production/decay modes.  Note also that
the position of this feature is given by (\ref{two-body}),
with $m_{\widetilde{\chi}_1^0}$ replaced by $m_{\widetilde{\chi}_2^0}$, 
which in this case is quite different from 
$m_{\widetilde{\chi}_4^0} - m_{\widetilde{\chi}_2^0} 
\doteq 100.2\, \hbox{GeV}$.  Thus care must be taken before assuming that
features in invariant mass plots correspond to --ino mass differences.

The designations in the last two paragraphs agree well with the
percentages given in Table \ref{tab:percents}, including the `stripe'
assignment above as well as the absence of a
$\widetilde{\chi}_3^0$-associated box or wedges in Figure \ref{pointA}.
The events lying outside the outer box in the Dalitz-like plot
are due at least in part to production modes including charginos.
This was confirmed in the HERWIG simulation by checking the identities of
the parent particles of the leptons in these outlying events.  In
addition, a sampling of such events were also found to have 
leptons from top-quark decays or lost leptons ({\it i.e.}, they were 
5 lepton events with one of the leptons being too soft to pass the
minimum $E_T$ cut or too close to the beam axis). 

\begin{figure}[t]
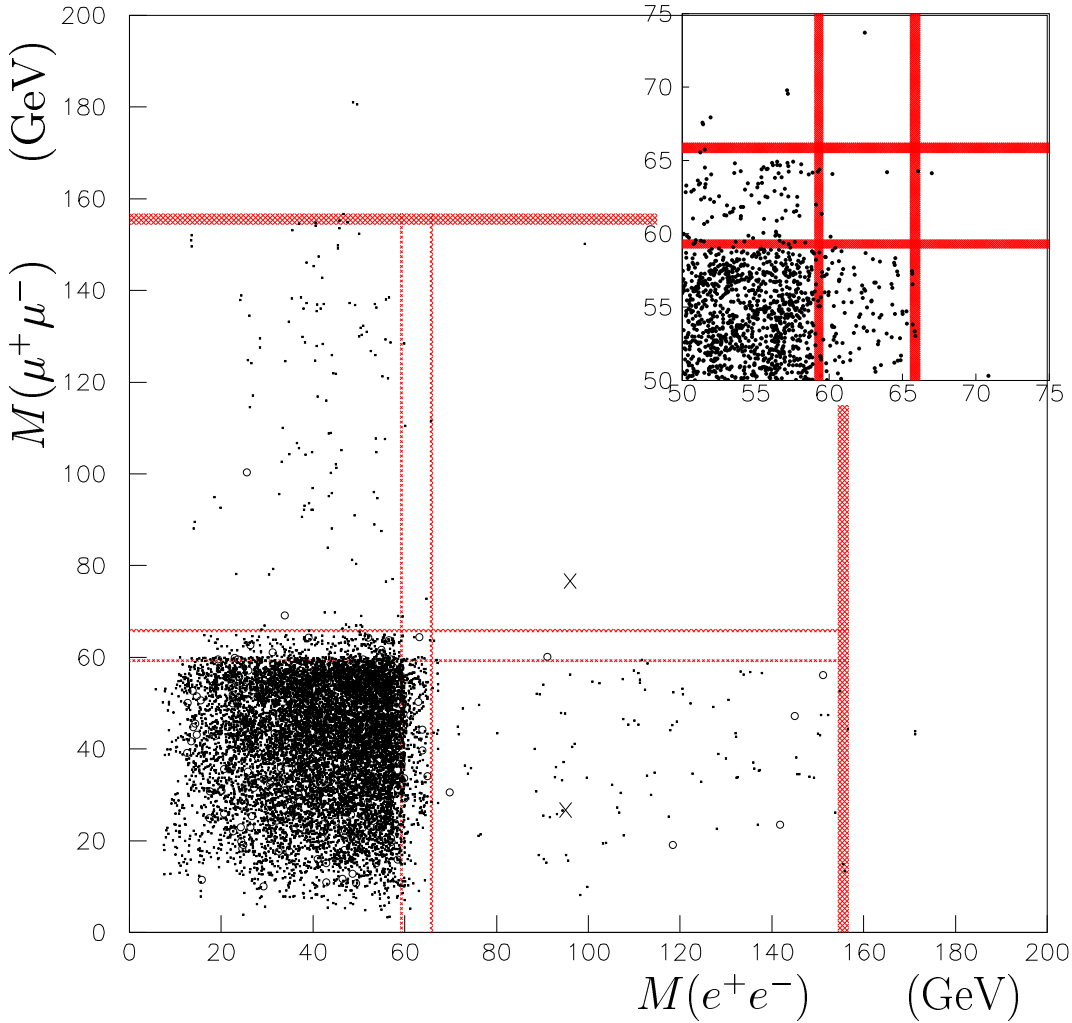

\vskip 3.1cm
\hspace{0.5cm}\dofig{14.0cm}{figure5.eps}
\vskip -9.25cm 
  \caption{{\small
$M(e^+e^-)$ {\it vs.} $\! M(\mu^+ \mu^-)$
Dalitz-like plot for MSSM Point $B$ assuming an integrated luminosity
of $30\, \hbox{fb}^{-1}$.
Signal events from gluino/squark production are denoted by dots,
SM background events by crosses and events from other SUSY processes
by open circles.
Shaded bands indicate kinematical endpoints expected based on ISASUSY.
The insert in the upper right corner zooms in on a critical region, and
the assumed integrated luminosity is raised to $60\, \hbox{fb}^{-1}$.
}}
\label{pointB}
\end{figure}

Figure \ref{pointB} for Point $B$ displays a somewhat
sparsely-populated wedge envelope matching pattern $B$ in the right-side
square of Figure \ref{boxes}.
The interior edges for the wedge are at ${\sim}60$-$65\, \hbox{GeV}$,
and event points taper off around $140$-$180\, \hbox{GeV}$.   
Inside this wedge is a much more densely-populated box with edges at 
roughly $60\, \hbox{GeV}$.
A second very short-legged wedge structure is also indicated, with
edges at $60\, \hbox{GeV}$ and $65\, \hbox{GeV}$.  More events from 
a longer run time would help to clarify the structure in the 
crucial 
($M(e^+ e^-),M(\mu^+ \mu^-)) =$ 
                        ($60$-$65\, \hbox{GeV},60$-$65\, \hbox{GeV}$)
region of the Dalitz-like plot (see insert in Figure \ref{pointB}).
The plot bespeaks of dominant 
$\widetilde{\chi}_i^0\widetilde{\chi}_i^0$ production with 
weaker contributions from 
$\widetilde{\chi}_i^0\widetilde{\chi}_j^0$ 
and $\widetilde{\chi}_i^0\widetilde{\chi}_k^0$
($i < j < k$) (the latter yielding the outer wedge envelope and the
former the short-legged wedge).  Note in the MSSM framework 
$i$, $j$ and $k$ must be $2$, $3$ and $4$.
The short, stubby wedge tells us
that two of the heavier --inos, presumably $\widetilde{\chi}_2^0$ and
$\widetilde{\chi}_3^0$, are quite close in mass.
This is in very good agreement with the predictions from 
Table \ref{tab:percents}:
a densely-populated $\widetilde{\chi}_2^0\widetilde{\chi}_2^0$ box  
and a short, stubby $\widetilde{\chi}_2^0\widetilde{\chi}_3^0$ wedge
(the $\widetilde{\chi}_3^0\widetilde{\chi}_3^0$ box is too sparsely
populated to be recognised --- again the
($M(e^+ e^-),M(\mu^+ \mu^-)) =$
                        ($60$-$65\, \hbox{GeV},60$-$65\, \hbox{GeV}$)
region of the Dalitz-like plot is seen to be crucial, with
more statistics desirable to clarify the situation.
Also, for this point in MSSM parameter space,
$m_{\widetilde{\chi}_3^0}$ is in fact rather close to
$m_{\widetilde{\chi}_2^0}$.

Shaded bands in the plot again show the expected locations of
hard edges based on the --ino and slepton mass spectrum obtained from
ISASUSY.  Though the $62.4\, \hbox{GeV}$
$\widetilde{\chi}_2^0$-$\widetilde{\chi}_1^0$
mass difference from ISASUSY roughly fits the position of the box edges, 
ISASUSY also reveals that the
$\widetilde{\chi}_2^0$ decays nearly always through an on-shell slepton,  
BR$(\widetilde{\chi}_2^0 \to \tilde{\ell}\ell \to
\widetilde{\chi}_1^0 + \ell^+\ell^-) = 0.999$, with the lighter
(predominantly right) and heavier (predominantly left) slepton
mass eigenstates contributing about equally.  Significantly,
the spoiler decay modes to sneutrinos only have a suprisingly 
low BR of only ${\sim}10^{-3}$.  
Applying (\ref{two-body}) using only the physical selectron masses 
(to match the HERWIG inputs) from Table \ref{tab:masses}
predicts edges at $58.5\, \hbox{GeV}$ and $59.0\, \hbox{GeV}$, confirming
that the inner box is from $\widetilde{\chi}_2^0\widetilde{\chi}_2^0$
production.  Again, $\widetilde{\chi}_3^0$ almost always decays via
on-shell sleptons to $\widetilde{\chi}_1^0$, but now $71.6$\% of the
decays are into sneutrino spoiler modes yielding no charged leptons.  
Application of (\ref{two-body}) now predicts endpoints at $64.8\,
\hbox{GeV}$ and $65.5\, \hbox{GeV}$, about $6\, \hbox{GeV}$ less than
$m_{\widetilde{\chi}_3^0} - m_{\widetilde{\chi}_1^0}$.

Note that gluino decays to $\widetilde{\chi}_4^0$ or
$\widetilde{\chi}_2^{\pm}$ are kinematically impossible and events
including a gluino decay to $\widetilde{\chi}_1^{\pm}$ cannot generate
$4\ell$ events.  Yet Table \ref{tab:percents} says that $3.0$\% of the 
$4\ell$ events are from $\widetilde{\chi}_2^0\widetilde{\chi}_4^0$
production and an outer $\widetilde{\chi}_2^0\widetilde{\chi}_4^0$ wedge
is clearly visible in the Dalitz-like plot.  This outer wedge must be
due solely to production of the heavier squarks which are heavy enough to
allow decays to $\widetilde{\chi}_4^0$ (the very small contribution from
$\widetilde{\chi}_4^0\widetilde{\chi}_4^0$ is insufficient
to generate the apparently-missing outer box).
The more massive $\widetilde{\chi}_4^0$ 
decays $59.4$\% of the time into sleptons ($26.6$\% of the time into 
charged sleptons and $32.8$\% of the time into sneutrinos).  The predicted
endpoints for the charged slepton decays from (\ref{two-body}) are
$153.0\, \hbox{GeV}$ and $155.4\, \hbox{GeV}$, basically
giving the outer ends of the wedge envelope over $80\, \hbox{GeV}$ below
the $\widetilde{\chi}_4^0$-$\widetilde{\chi}_1^0$ mass difference.
Again, some if not all of the events lying outside these bounds
come from processes involving charginos.

\begin{figure}[t]
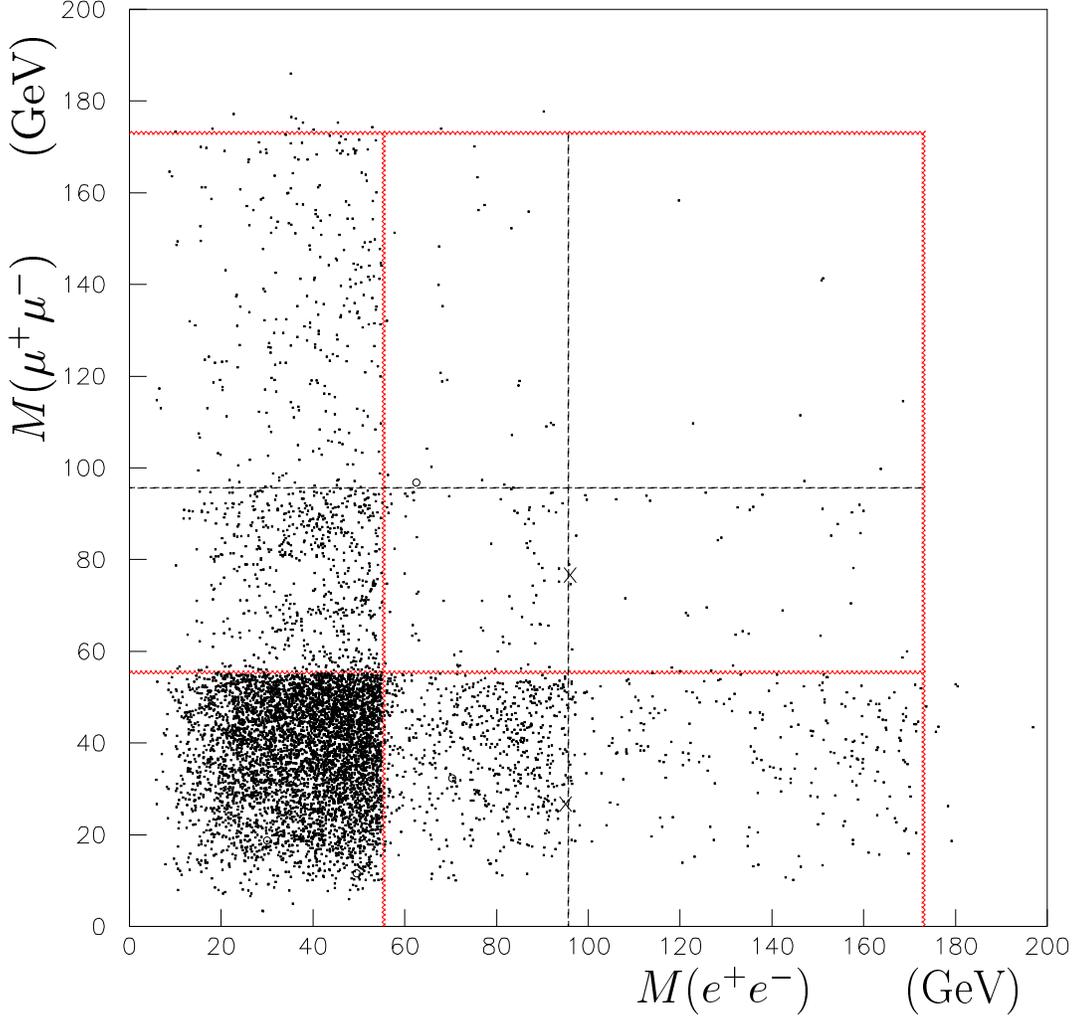

\vskip 3.1cm
\hspace{0.5cm}\dofig{14.0cm}{figure6.eps}
\vskip -9.25cm 
\caption{{\small
$M(e^+e^-)$ {\it vs.} $\! M(\mu^+ \mu^-)$
Dalitz-like plot for MSSM Point $C$ assuming an integrated luminosity 
of $30\, \hbox{fb}^{-1}$.
Signal events from gluino/squark production are denoted by dots,
SM background events by crosses and events from other SUSY processes
by open circles.
Shaded bands indicate kinematical endpoints expected based on ISASUSY.
The band marking the endpoint for
$\widetilde{\chi}^0_3 \rightarrow \tilde{\ell}^{\pm} \ell^{\mp}
\rightarrow \ell^+ \ell^-$ decays is quite narrow, and is drawn as a 
dashed line in this plot.
}}
\label{pointC}
\end{figure}

Lastly, Figure \ref{pointC} shows the Dalitz-like plot obtained 
for MSSM parameter point $C$.  A wedge extending out to 
${\sim}175$-$180\, \hbox{GeV}$ is readily seen.
The inner edges of this wedge are at ${\sim}55\, \hbox{GeV}$.
Inside of this wedge is a shorter wedge with the same inner edges
terminating at ${\sim}95\, \hbox{GeV}$, and in the corner a
densely-populated box.  Structures outside this wedge are more
difficult to discern with this number of events:  a box with
edges at $95\, \hbox{GeV}$ is somewhat clear while a wedge with
inner edges at ${\sim}95\, \hbox{GeV}$ extending out to ends at 
${\sim}175$-$180\, \hbox{GeV}$ may be barely discernible.
Thus this plot could be classified as either pattern $B$ or pattern $C$
according to the nomenclature introduced in the right-side square of
Figure \ref{boxes}.
The features exhibited suggest fairly dominant
$\widetilde{\chi}_i^0\widetilde{\chi}_i^0$ production, but with 
significant contributions from 
$\widetilde{\chi}_i^0\widetilde{\chi}_j^0$
and $\widetilde{\chi}_i^0\widetilde{\chi}_k^0$
($i < j < k$), and with lesser but still detectable 
contributions from 
$\widetilde{\chi}_j^0\widetilde{\chi}_j^0$ and
$\widetilde{\chi}_j^0\widetilde{\chi}_k^0$. 
(Again, in the MSSM, $i$, $j$ and $k$ must be $2$, $3$ and $4$.)

ISASUSY numbers for the --ino and slepton mass spectrum again
yield the shaded bands and dashed lines showing the expected locations 
of hard edges.
$\widetilde{\chi}_2^0$ virtually always decays via on-shell sleptons
to $\widetilde{\chi}_1^0$, $69.5$\% of the time through charged 
sleptons and $30.5$\% of the time through sneutrinos (contrast this 
with $\widetilde{\chi}_2^0$ decays at MSSM parameter point $B$).
Again applying (\ref{two-body}) using only the physical
selectron masses (to match the HERWIG inputs) from Table \ref{tab:masses}
leads to predicted edges at $54.4\, \hbox{GeV}$ and $55.3\, \hbox{GeV}$,
corroborating that the inner box is from
$\widetilde{\chi}_2^0\widetilde{\chi}_2^0$ production. 
Note that use of more correct physical smuon masses incorporating 
left-right sfermion mixing would significantly widen the horizontal
shaded bands in Figure \ref{pointC}.

Comparing the wedges in Figure \ref{pointC} with the one in
Figure \ref{pointB}, we conclude that in this case the --ino masses
are not so close together.   
$\widetilde{\chi}_3^0$ decays via on-shell sleptons $77.6$\% of the time
($17.7$\% via charged sleptons and $59.9$\% via sneutrinos), the rest of
the time decaying via an on-shell $Z^0$ ($21.6$\%) or a
$\widetilde{\chi}_1^{\pm}$ ($0.7$\%) or a $\widetilde{\chi}_2^0$
($<0.1$\%).  The charged-slepton-mediated $\widetilde{\chi}_3^0$ 
decays should have endpoints at $95.8\, \hbox{GeV}$ and 
$95.9\, \hbox{GeV}$ (in this case the two-body endpoint is 
nearly equal to the $\widetilde{\chi}_3^0$-$\widetilde{\chi}_1^0$
mass difference).
The $\widetilde{\chi}_3^0$ decays via $Z^0$ lead to a band at 
$91.2\, \hbox{GeV}$ also faintly visible in Figure \ref{pointC}.
As with MSSM parameter point $B$, the sneutrino spoiler modes 
are much stronger (considerably stronger) in 
$\widetilde{\chi}_3^0$ ($\widetilde{\chi}_4^0$) decays
than for $\widetilde{\chi}_2^0$ decays, suppressing contributions 
from the former to the Dalitz-like plot relative to the latter.
Decays of $\widetilde{\chi}_4^0$ via on-shell charged sleptons 
(which occurs $26.9$\% of the time, compared to $43.7$\% of the decays 
being via sneutrinos) will result in edges at $172.3\, \hbox{GeV}$ and
$173.0\, \hbox{GeV}$ ($10\, \hbox{GeV}$ or so below
$m_{\widetilde{\chi}_4^0} - m_{\widetilde{\chi}_1^0}$).  
$\widetilde{\chi}_4^0$ also decays\footnote{ There
are also a smattering of other $\widetilde{\chi}_4^0$ decay modes:  
to staus $1.4$\% of the time, to $\widetilde{\nu}_{\tau}$ $3.8$\%,
$\rightarrow \widetilde{\chi}_2^0(\widetilde{\chi}_1^0) + h^0$
$1.2$\%($0.1$\%), and 
$\rightarrow \widetilde{\chi}_2^0(\widetilde{\chi}_1^0) + Z^0$
$0.2$\%($0.1$\%).  These could only contribute a very small fraction 
of the events.} $22.3$\% of the time into
$\widetilde{\chi}_1^{\pm}W^{\mp}$ which can yield aberrant
events not anticipated in the neutralinos-only framework followed here.
The variation in the widths of the shaded bands due to
decays occurring through the two different same-flavour sleptons, 
which are $4.06$, $0.44$, and $3.24\, \hbox{GeV}$, can readily  
be understood from the variation of (\ref{two-body}):
\begin{equation}
\label{two-body-deriv}
\frac{\partial M(\ell^+\ell^-)}{\partial{m_{\tilde{\ell}}}^2}
= \frac{\Delta {m_{\tilde{\ell}}}^2}{2 M(\ell^+\ell^-)}
\left( \frac{{m_{\widetilde{\chi}_1^0}}^2 {m_{\widetilde{\chi}_i^0}}^2 }
            {{m_{\tilde{\ell}}}^4}-1
           \right) \, .
\end{equation}
In this case the $\widetilde{\chi}_2^0 \rightarrow \widetilde{\chi}_1^0$
 and $\widetilde{\chi}_4^0 \rightarrow \widetilde{\chi}_1^0$ bands have 
similar widths, as this is inversely proportional to the endpoint yet
partially compensated for by the factor in parentheses for much heavier
$\widetilde{\chi}_4^0$; this factor is however very small for
intermediate-mass $\widetilde{\chi}_3^0$,
where
$ {m_{\widetilde{\chi}_1^0}}^2 {m_{\widetilde{\chi}_3^0}}^2
            /{m_{\tilde{\ell}}}^4 \approx 1$, hence the relatively thin
$\widetilde{\chi}_3^0 \rightarrow \widetilde{\chi}_1^0$ band.

Summarizing, the predicted endpoints from charged-slepton-mediated
decays of $\widetilde{\chi}_3^0$ and $\widetilde{\chi}_4^0$ affirm that 
the shorter (longer) wedge is from 
$\widetilde{\chi}_2^0\widetilde{\chi}_3^0$
($\widetilde{\chi}_2^0\widetilde{\chi}_4^0$)
production.  The more faintly discernible box with edges at  
${\sim}95\, \hbox{GeV}$ is attributed to
$\widetilde{\chi}_3^0\widetilde{\chi}_3^0$ production, the
even more faint wedge of which this box is the corner
is from $\widetilde{\chi}_3^0\widetilde{\chi}_4^0$ production,
and the few events in the upper-left corner of the plot are 
presumably from $\widetilde{\chi}_4^0\widetilde{\chi}_4^0$ 
production. 
The relative percentages of $4\ell$ events given in 
Table \ref{tab:percents} agrees fairly well with the densities 
of points seen in the associated features in Figure \ref{pointC}.

It is interesting to see how effectively the
relative --ino pair contributions can be extracted from the 
Dalitz-like plot.  
Assuming some knowledge of the dilepton invariant mass
distribution, an estimate of the ratio of the
different --ino pair production rates
(stemming from the gluino/squark BRs to the different --inos) 
is obtainable from counting the total number of points in each 
region of the Dalitz plot and then taking the ratio.
Approximating the distributions as being exactly triangular
(see $\!\!\!\!\!$ \cite{PaigeII}),
and taking the endpoint locations noted in the preceding paragraphs
as $55\, \hbox{GeV}$ , $96\, \hbox{GeV}$ and  $173\, \hbox{GeV}$,
the following rate comparison can be extracted\footnote{Calculational
details are relegated to an appendix.}:
\begin{equation}
r_{22} : r_{23} : r_{24} : r_{33} : r_{34} : r_{44}
= 431 : 118 : 59 : 15.5 : 9.4 : 1 \;\;\; ,
\end{equation}
where $r_{ij}$ is the rate from
$\widetilde{\chi}_i^0 \widetilde{\chi}_j^0$ production, or,
considering just the the three wedges,
$r_{23} : r_{24} : r_{34} = 12.3 : 6.3 : 1 \, $.
Compare these values to the results obtained earlier from 
Table \ref{tab:percents}
\begin{equation}
r_{22} : r_{33} : r_{44} = 131.5 : 1.3 : 1 \;\; \hbox{and} \;\;
 r_{23} : r_{24} : r_{34} = 10.2 : 9.6 : 1 \;\; .
\end{equation}
Little more than crude agreement is discernible; bear in mind though
that, as noted earlier, discrepancies may be reasonably expected in
comparing all-inclusive $4\ell$ rates with $e^+e^-{\mu}^+{\mu}^-$ rates
after cuts.  The assumption of strictly triangular population profiles
is also certainly somewhat inaccurate.  And, at least with modest
statistics ({\ie}, with only results from the first year or two of 
running for the LHC), there will be significant imprecision in pinpointing
the locations of the endpoints (the main source of uncertainty in the 
calculation if the triangular distribution assumption is viable).
One factor that is {\em not} an important concern at this MSSM
parameter
point is contamination from chargino-related events; but, said
contamination could skew such a calculation at other MSSM points (as, for
instance, Point $A$).  

If $r_{ij} = 2 r_i r_j$ is now assumed to be valid
(note though the afore-mentioned serious caveats to this assumption),
the relative individual production times leptonic BRs are obtainable:
$ r_{2} : r_{3} : r_{4} = 12.3 : 1.6 : 1 \; $
(using $r_{23}$, $r_{34}$ and $r_{22}$ or $r_{24}$ as inputs)
or
$ r_{2} : r_{3} : r_{4} = 12.6 : 4.7 : 1 \; $ 
(using $r_{23}$, $r_{34}$ and $r_{44}$ as inputs).
The extent to which these two results disagree
could (as before for the inclusive $4\ell$ results)
be interpreted as implying significant contributions from
squark-production (though the inaccuracy of the triangular 
distribution assumption may also be a factor).  
Recall inclusive $4\ell$ results were
$ r_2 : r_3 : r_4 = 19.00 : 1.04 : 1 $
(using $r_{23}$, $r_{34}$ and $r_{24}$ as inputs)
and 
$ r_2 : r_3 : r_4 = 11.47 : 1.16 : 1 $
(using $r_{22}$, $r_{33}$ and $r_{44}$ as inputs).
Apparently, dynamical information from the densities of events
in the Dalitz-like plot's various geometrical components may be more
difficult to extract than the kinematical information contained in the 
location of the hard edges.  However, more sophisticated statistical
analyses may be expected to yield better results.

\begin{figure}[t]
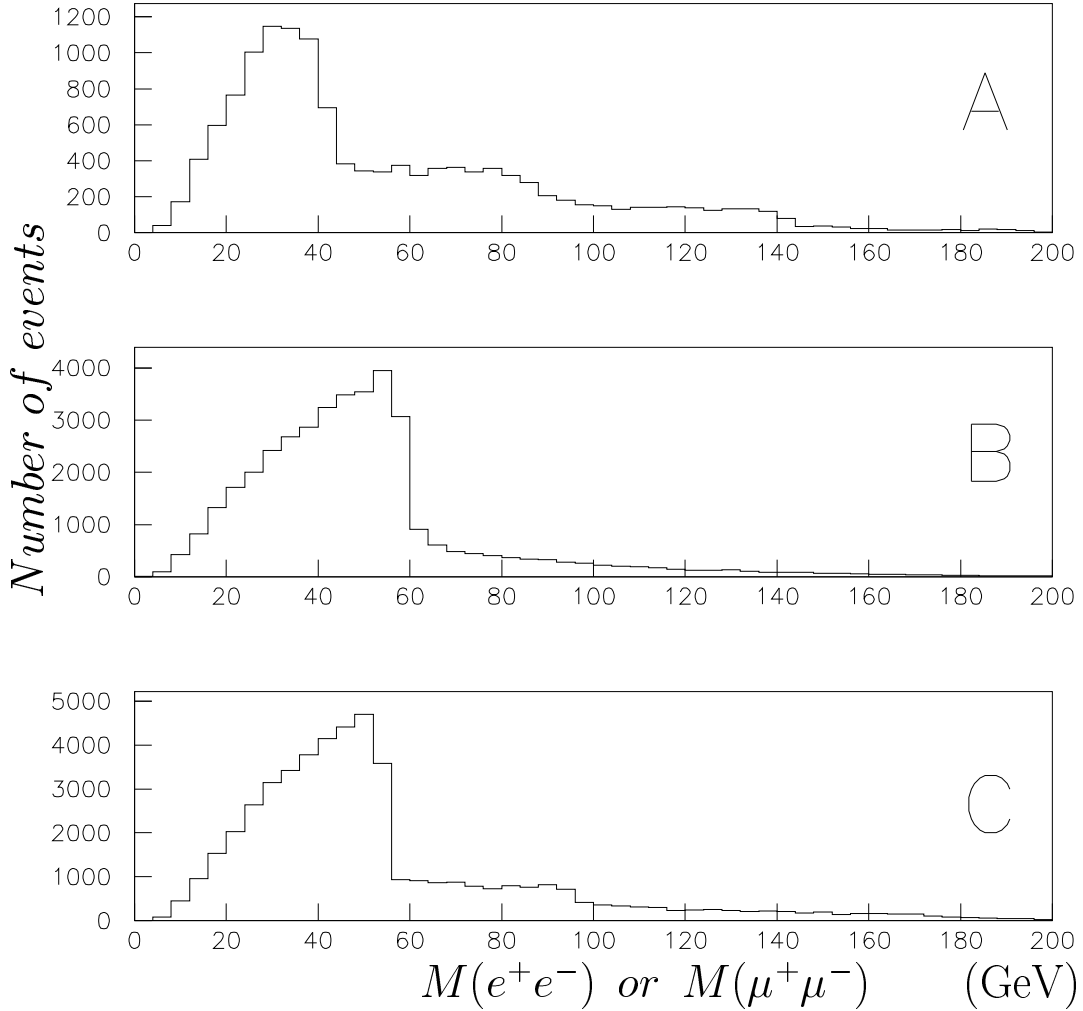

\vskip 2.8cm
\hspace{0.7cm}\dofig{14.0cm}{figure7.eps}
\vskip -9.25cm 
  \caption{{\small
One-dimensional projection of 
Figures \ref{pointA}-\ref{pointC} 
for MSSM Points $A$, $B$ and
$C$ assuming an integrated luminosity 
of $30\, \hbox{fb}^{-1}$.  Information about the decay topology is lost in
such projections.
}}
\label{pointABC-1}
\end{figure}

Next contrast the information apparent in the Dalitz-like plots with 
that readily obtainable from the more traditional one-dimensional 
projections shown in Figure \ref{pointABC-1}.  
Notice how similar the results for Points $B$ and $C$ appear in
Figure \ref{pointABC-1}, while Figure \ref{pointB} and
Figure \ref{pointC} are quite different.
Note also in this case the sharp drops observed would only be sufficient
to identify which --inos are being produced,
{\em not} which --ino {\em pairs} are being produced.

\section*{Concluding Remarks}

Production of pairs of new heavy particle states $X_i X_j$
at hadron colliders has been studied emphasising the simple topological
forms expected in certain two-dimensional Dalitz-like plots.  
It is assumed the heavy particles decay into pairs of SM particles 
(with the possible addition of substantial amounts of missing energy),
yielding final states of the form
$x \bar{x} \, + \, y \bar{y} \,\, (+ \,  \slashchar{E}$), 
where in this work $x$ and $y$ are taken to be distinct SM fermions.
Given a sufficient number of events, the observed topology
(a `box' or a `wedge') clearly indicates whether or not $X_i$ and $X_j$
are identical particles.  When simultaneous production of more than
one pair of new particles is possible within a model, a more extensive set
of topologies constructed from boxes and wedges (possibly with overlayed
`stripes') is obtained.
A likelihood function indicating how well the a set of data points fits 
each possibility can be readily constructed if visual inspection
does not suffice.  The particular set of shapes the data sample
should be thus compared to is of course model dependent.

Though we wish to stress the general applicability of this 
technique to a fairly wide range of beyond-the-SM scenarios, 
application to $R$-parity conserving SUSY models 
readily springs to mind.  Thus the pair production of heavy MSSM 
neutralinos (excluding the lightest one, the LSP), with the subsequent 
decay of each --ino into a pair of leptons to aid identification, has 
been examined in detail.
Here a fairly sizable number of distinct topological shapes 
is obtainable.  This work then further specialises to --ino pairs
produced in gluino/squark decays, most likely to be the dominant mode of
--ino production at the LHC --- if gluinos and/or squarks are relatively 
light.  The number of possible topologies may be substantially reduced
when this is the production mechanism compared to the EW production
mechanisms which contain an --ino--ino-$S$ vertex if squark production
does not re-introduce complexity.  This was examined in some detail
including possible tests of simulation results that may indicate the
significance of squark production (and distinguish it from gluino pair
production).  Neutralino results thus obtained might be compared to those 
from charge asymmetries possible in samples of like-sign dilepton events 
from chargino pair production $\!\!\!\!\!$ \cite{BaerTatHigh}.

The `hard edges' seen in a Dalitz-like plot yield information on the
--ino mass differences as well as the identities of --inos
participating in the decays (though it should be emphasised that the 
endpoints certainly need not equal the mass difference of two --inos
if on-mass-shell sleptons are involved in the decay chains), while 
comparing the relative densities of regions populated by different 
$X_i X_j$ production channels or combinations of channels has the 
potential to provide information on the relative production cross
sections times leptonic BRs of these channels.
We found simulation results from HERWIG for three distinctive points in 
MSSM parameter space (including cuts that nearly eliminate the 
backgrounds and a realistic detector simulation) clearly closely tract
the partial results we obtained at these points by `hand'-calculations
based on the ISASUSY inputs.  It is apparent from the Dalitz-like
plots shown that this includes a substantial amount of information not 
available from a one-dimensional plot that just lumps together $e^+e^-$ and 
${\mu}^+{\mu}^-$ invariant masses.  Upcoming consideration of
--ino pair production via heavy MSSM Higgs bosons $\!\!\!\!$ \cite{inPrep}, 
which also can have quite substantial rates in favourable regions of MSSM
parameter space, will further expand the extra information
obtainable from the two-dimensional Dalitz-like plot and thus 
should prove very exciting.
Of course such analyses only incorporate
mixed leptonic decays\footnote{Unless angular
correlations between leptons can be exploited to say how 
four same-flavour leptons should be arranged into two pairs
without prejudicing the distribution.  Or one could just plot all
possible opposite-sign pair combinations, such plots may at least be
distinguishable for different --ino pairs.} --- ($e^+e^-{\mu}^+{\mu}^-$
events, but not $e^+e^-e^+e^-$ or ${\mu}^+{\mu}^-{\mu}^+{\mu}^-$ events).

We also note that it is possible to make lego-style 3-dimensional
plots with $M(x\bar{x})$ and $M(y\bar{y})$ along two axes and the binned
number of events along a third axis.  Figures obtained in this way
were not found particularly illuminating for the specific processes
and MSSM parameter points studied here (and with the modest amount of
integrated luminosity assumed), but may be more useful in other studies. 

How far this method can go toward aiding
reconstruction of the --ino mass spectrum will depend on particulars
of the point in the MSSM parameter space nature chooses, but clearly
very significant information may be extracted.  Given that an
$e^+e^-$ linear collider with a centre-of-mass energy beyond that of 
LEP 2 is not expected for some time, it is crucial to seek the
optimal methods for disentangling the --inos produced at the LHC.
Further, information on the heavier --ino states may prove crucial in 
deciding the reach of a future linear collider to perform the 
more precise measurements surely required.

\section*{Acknowledgements}
 We thank Y.N. Gao and X. Tata for useful comments.
This work was supported by Tsinghua University.
SM is partially supported by UK-PPARC.

\section*{Appendix}
The schematic Dalitz-like plot shown in the right square of Figure
\ref{boxes} is a collection of 6 observables 
(there labeled as regions $\alpha, \beta, ...\kappa$) from which
the production times leptonic BR values for the various --ino pairs,
$r_{ij}$ ($i,j=2,3,4$) may be extracted.  
First, a triangular distribution of events is assumed for each 
--ino --ino mode:
\begin{equation}
r_{ij} = {\cal K} \int dx \int dy \; x \, y
\end{equation}
where ${\cal K}$ is a normalization constant that will drop out of the
calculation. 
Now each region of the Dalitz plot contains events attributable to one 
or more of the modes $r_{ij}$. The six different regions therefore
correspond to different 
combinations of the $r_{ij}$; which may be written as
\begin{equation}
v_1 = {\cal M} v_2 \; ,
\end{equation}
with vectors $v_1 = \left( \alpha, 2 \beta, 2 \gamma, \delta, 
2 \eta, \kappa \right) $ and
$v_2 = \left( r_{22}, r_{23},r_{24},r_{33},r_{34},r_{44} \right) $, 
and the matrix 
\begin{equation}
{ \cal M}  = \left(
\begin{array}{c c c c c c}
1 & a & b & c & d & e \\
0 & f & g & h & i & j \\
0 & 0 & k & 0 & l & m \\
0 & 0 & 0 & n & o & p \\
0 & 0 & 0 & 0 & q & r \\
0 & 0 & 0 & 0 & 0 & s \\
\end{array}
\right)
\end{equation}
where $a,b,c,...s$ are numbers between 0 and 1 which represent the
fraction 
of events from $r_{ij}$ in a particular region; for example, 
$s= \frac{\int^{E_2}_{E_1} dx \int^{E_2}_{E_1} dy \; x \, y}
{\int^{E_2}_{0} dx \int^{E_2}_0 dy \; x \, y} 
= \left[ 1- \left(\frac{E_1}{E_2}\right)^2 \right]^2$
with $E_{0,1,2}$ being the three kinematical endpoints in the figure.
Elements in each column of the matrix ${\cal M}$ must sum to unity.
Now defining $x = \left( \frac{E_0}{E_1} \right)^2$, 
$y = \left( \frac{E_0}{E_2} \right)^2$ and
$z = \left( \frac{E_1}{E_2} \right)^2$ yields
\newline
\begin{tabular}{cccc}  
$a = x$
&
$b = y$
&
$c = x^2$
&
$d = x y$
 \\
$e = y^2$
&
$f = 1-x$
&
$g = z-y$
&
$h = 2 x (1-x)$
 \\
$i = 2 y (1-x)$
&
$j = 2 y (z-y)$
&
$k = 1 - z$
&
$l = x (1-z)$
 \\
$m = 2 y (1-z)$
&
$n = (1-x)^2$
&
$o = (1-x)(z-y)$
&
$p = (z-y)^2$
 \\
$q = (1-x)(1-z)$
&
$r = 2 (1-z)(z-y)$
&
$s = (1-z)^2$
&
\end{tabular}
\vskip 0.2cm
\noindent
The linear system of equations is now easily solved for the individual 
rates $r_{ij}$.

\def\pr#1 #2 #3 { {\rm Phys. Rev.}            {#1}, #3 (#2)}
\def\prd#1 #2 #3{ {\rm Phys. Rev. D}          {#1}, #3 (#2)}
\def\prl#1 #2 #3{ {\rm Phys. Rev. Lett.}      {#1}, #3 (#2)}
\def\plb#1 #2 #3{ {\rm Phys. Lett. B}         {#1}, #3 (#2)}
\def\npb#1 #2 #3{ {\rm Nucl. Phys. B}         {#1}, #3 (#2)}
\def\prp#1 #2 #3{ {\rm Phys. Rep.}            {#1}, #3 (#2)}
\def\zpc#1 #2 #3{ {\rm Z. Phys. C}            {#1}, #3 (#2)}
\def\epjc#1 #2 #3{ {\rm Eur. Phys. J. C}      {#1}, #3 (#2)}
\def\mpl#1 #2 #3{ {\rm Mod. Phys. Lett. A}    {#1}, #3 (#2)}
\def\ijmp#1 #2 #3{{\rm Int. J. Mod. Phys. A}  {#1}, #3 (#2)}
\def\ptp#1 #2 #3{ {\rm Prog. Theor. Phys.}    {#1}, #3 (#2)}
\def\jhep#1 #2 #3{ {\rm J. High Energy Phys.} {#1}, #3 (#2)}
\def\jphg#1 #2 #3{ {\rm J. Phys. G}           {#1}, #3 (#2)}

\pagebreak

\end{document}